\documentclass[11pt,aps,prd,preprint,groupedaddress,tightenlines,nofootinbib,floatfix]{revtex4}
\usepackage[dvipdfm]{graphicx}

\usepackage{amssymb}
\usepackage{amsmath}
\usepackage{epstopdf}
\usepackage{booktabs}

\usepackage{subfigure}
\makeatletter
\newcommand{\figcaption}[1]{\def\@captype{figure}\caption{#1}}
\newcommand{\tblcaption}[1]{\def\@captype{table}\caption{#1}}
\makeatother

\def\simge{\mathrel{%
       \rlap{\raise 0.511ex \hbox{$>$}}{\lower 0.511ex \hbox{$\sim$}}}}
\def\simle{\mathrel{
       \rlap{\raise 0.511ex \hbox{$<$}}{\lower 0.511ex \hbox{$\sim$}}}}

\begin{document}

\title{Histograms in heavy-quark QCD at finite temperature and density}
\author{H.~Saito$^{1}$\footnote{Present address: 
NIC, DESY Zeuthen, Platanenallee 6, 15738 Zeuthen, Germany}, 
S.~Ejiri$^{2}$, S.~Aoki$^{1,3}$\footnote{Present address: 
Yukawa Institute for Theoretical Physics, Kyoto University,
Kitashirakawa Oiwakecho, Sakyo-ku, Kyoto 606-8502, Japan}, 
K.~Kanaya$^{1}$, Y.~Nakagawa$^{2}$, H.~Ohno$^{4}$, K.~Okuno$^{2}$, 
and T.~Umeda$^{5}$ \\
(WHOT-QCD Collaboration)
}
\affiliation{$^1$Graduate School of Pure and Applied Sciences, University of Tsukuba, Tsukuba, Ibaraki 305-8571, Japan\\
$^2$Graduate School of Science and Technology, Niigata University, Niigata 950-2181, Japan\\
$^3$Center for Computational Sciences, University of Tsukuba, Tsukuba, Ibaraki 305-8577, Japan\\
$^4$Physics Department, Brookhaven National Laboratory, Upton, New York 11973, USA\\
$^5$Graduate School of Education, Hiroshima University, Hiroshima 739-8524, Japan
}
\date{January 14, 2014}

\begin{abstract}
We study the phase structure of lattice QCD with heavy quarks at finite temperature and density by a histogram method. 
We determine the location of the critical point at which the first-order deconfining transition in the heavy-quark limit turns into a crossover at intermediate quark masses through a change of the shape of the histogram under variation of coupling parameters.
We estimate the effect of the complex phase factor which causes the sign problem at finite density, and show that, in heavy-quark QCD, the effect is small around the critical point.
We determine the critical surface in $2+1$ flavor QCD in the heavy-quark region at all values of the chemical potential $\mu$ including $\mu=\infty$.
\end{abstract}

\maketitle

\section{Introduction}
\label{introduction}

Quantum chromodynamics (QCD) has a rich phase structure as a function of temperature $T$, quark chemical potential $\mu$, and quark mass $m_q$  \cite{review1}.
Determination of it, in particular the location of the transition between the confined phase at low $T$ and $\mu$ and the deconfined phase at high $T$, plays important roles in a study of early evolutions in the Universe.
From lattice QCD simulations, the transition is considered to be an analytic crossover at low $\mu$ around the physical point of quark masses but is expected to turn into a first-order transition when we increase $\mu$ or vary the quark masses.
Identification of the critical point where the crossover turns into the first-order transition is also important in understanding the nature of the quark matter created in high-energy heavy-ion collision experiments \cite{review2}.

When we increase $\mu$ in a lattice simulation of QCD, the sign problem becomes severe. 
An extremely high statistics is required to obtain statistically significant results from Monte Carlo (MC) simulations when $\mu$ is not small enough.
We need to combine various techniques developed at small $\mu$ to extend the range of calculability toward larger $\mu$. 
One of the goals of these studies is to find a decisive evidence of the first-order transition by a MC study to locate the critical point.

To achieve this goal, we are developing a histogram method in which histograms are used to identify a first-order signal 
\cite{Ejiri:2007ga,PRD84_054502,WHOT_PTEP12,Ejiri2012}. 
Our histogram method can be viewed as a variant of the spectral density methods \cite{Gocksch88,Azcoiti90,Nishimura02,Fodor03}.

When the transition is of first order, we expect multiple peaks in the histogram for observables that are sensitive to the phase, such as the energy density and chiral order parameter. 
We can thus detect the on-set of a first-order transition through a deformation of the histogram.
To do this, however, we need statistically reliable data on the histogram over a wide range of expectation values. 
With a single MC simulation, a correct evaluation of a multiple-peak histogram at a first-order transition point requires many flip-flops among different phases, that is computationally demanding with dynamical quarks.
This problem is remedied by introducing the reweighting technique \cite{McDonaldSinger,Ferrenberg:1988yz,FS89},
with which we can combine simulations at different points in the parameter space, and thus with slightly different support of expectation values, to obtain a reliable histogram over a wide range of expectation values \cite{Ejiri:2007ga}.

In a previous paper \cite{PRD84_054502}, we adopted the histogram method to investigate the quark mass dependence of the QCD phase transition in the heavy quark region at $\mu=0$ and have shown that the histogram method is powerful in determining the phase structure: 
Combining a MC simulation in the heavy quark limit of QCD [SU(3) Yang-Mills theory] and the hopping parameter expansion, 
the first-order deconfining  transition of QCD in the heavy quark limit is shown to become weaker when we decrease the quark mass from infinity.
The location of the critical point where the first-order transition turns into a crossover \cite{Ukawa,DeTar,Alexandrou99,Fromm12} was computed using various observables. 
Extending the results to the case of $2+1$ flavor QCD, we thus determined the critical line separating the first order and crossover regions around the top-right corner of the Columbia plot \cite{PRD84_054502}.

In this study, we introduce chemical potential $\mu$ to the study of heavy quark QCD. 
The critical line in $2+1$ flavor QCD becomes a critical surface in the parameter space including $\mu$.
Although the critical surface in the heavy-quark region is different from that in the light-quark region\cite{crtpt,dFP03} and thus is not directly relevant to the transition in the real world, 
the former provides us with a good testing ground of the method, since the computational burden is much lighter.
We show that, to the leading order of the hopping parameter expansion, the complex phase of the quark determinant is introduced though the imaginary part of the Polyakov line.
By studying an effective potential of the Polyakov line, we find that the $\mu$-dependence of the critical surface is simple in the phase-quenched theory. 
We then take account of the effects of the complex phase by estimating the phase average using the cumulant expansion method \cite{Ejiri:2007ga,whot10} to determine the critical surface in $2+1$ flavor QCD at finite density.

This paper is organized as follows. 
We introduce the histogram method and its combination with a multi-point reweighing method in Sec.~\ref{sec:method}.
QCD in the heavy-quark region is discussed to the leading order of the hopping parameter expansion in Sec.~\ref{sec:heavy}.
In Sec.~\ref{sec:pldist0}, we study the histogram and its effective potential for the Polyakov line, as well as those for the plaquette, first at $\mu=0$. 
We find that both give a consistent result for the critical point.
In Sec.~\ref{sec:poldis}, we estimate the effect of the complex phase on the Polyakov line histogram at finite chemical potential.
We show that the effect is small around the critical point and the location of the critical point is well approximated by that in the phase-quenched theory.
We determine the critical surface in $2+1$ flavor QCD.
In Sec.~\ref{sec:popot},  we calculate the double-histogram simultaneously for the Polyakov line and plaquette.
By drawing the curves on which the first derivatives of the effective potential vanish and by tracing their intersection points, we study the fate of the first order transition.
The phase diagram thus obtained agrees with those obtained by the Polyakov line histogram or the plaquette histogram alone. 
We summarize our conclusions in Sec.~\ref{sec:conclusion}.
Appendix~\ref{sec:range} is added to show that the effect of the next-to-leading terms of the hopping parameter expansion is small on the location of the critical point.

\section{Histogram method}
\label{sec:method}

We study QCD with $N_{\rm f}$ flavors of Wilson-type quarks. 
We write the gauge action as 
\begin{eqnarray}
S_g = -6 N_{\rm site} \,\beta \, \hat{P},
\label{eq:Sg}
\end{eqnarray} 
where 
$\beta = 6/g^2$ is the gauge coupling parameter, 
$N_{\rm site} = N_s^3 \times N_t$ is the lattice volume,
and $\hat{P}$ is the (generalized) plaquette. 
For the case of the standard plaquette gauge action which we study in the next section, 
\begin{eqnarray}
\hat{P}= \frac{1}{18 N_{\rm site}} \displaystyle \sum_{x,\mu < \nu} 
 {\rm Re \ Tr} \left[ U_{x,\mu} U_{x+\hat{\mu},\nu}
U^{\dagger}_{x+\hat{\nu},\mu} U^{\dagger}_{x,\nu} \right] .
\label{eq:SgP}
\end{eqnarray} 
We write the quark action as
\begin{eqnarray}
S_q = \sum_{f=1}^{N_{\rm f}} \sum_{x,y} \bar{\psi}_x^{(f)} \, M_{xy} (\kappa_f,\mu_f) \, \psi_y^{(f)} ,
\label{eq:Sq}
\end{eqnarray} 
where $M_{xy}$ is the quark kernel
and $\kappa_f$ and  $\mu_f$ are the hopping parameter and chemical potential for the $f$th flavor, respectively. 
We denote $\vec\kappa = (\kappa_1,\cdots,\kappa_{N_{\rm f}})$ and  $\vec\mu = (\mu_1,\cdots,\mu_{N_{\rm f}})$ for the sets of hopping parameters and chemical potentials.
In this study, we assume that $M_{xy}$ is independent of $\beta$.
\footnote{For the cases of $\beta$-dependent $M$, see discussions in Sec.~6.5 of Ref.~\cite{WHOT_PTEP12}.}

\subsection{Histogram and effective potential}
\label{sec:histogram}

We define the histogram for a set of physical quantities 
$X = (X_1,X_2,\cdots)$ as 
\begin{eqnarray}
w(X; \beta, \vec\kappa, \vec\mu) 
&=& \int {\cal D} U {\cal D} \psi {\cal D} \bar{\psi} \ \prod_i \delta(X_i - \hat{X}_i) \ e^{- S_q - S_g} \nonumber \\
&=& \int {\cal D} U \ \prod_i \delta(X_i - \hat{X}_i) \ 
e^{-S_g}\ \prod_{f=1}^{N_{\rm f}} \det M(\kappa_f, \mu_f) .
\label{eq:dist}
\end{eqnarray}
where $\hat{X} = (\hat{X}_1,\hat{X}_2,\cdots)$ is the operators for $X$. 
Then the partition function is given by
\begin{equation}
Z(\beta,\vec\kappa,\vec\mu)  = \int\! w(X;\beta,\vec\kappa,\vec\mu) \, dX
\end{equation}
with $dX = \prod_i dX_i$, 
and the probability distribution function of $X$ is given by $Z^{-1} w(X;\beta,\vec\kappa,\vec\mu)$. 
The expectation value of an operator ${\cal O} [\hat{X}]$ which is written in terms of $\hat{X}$ is calculated by 
\begin{eqnarray}
\langle {\cal O }[\hat{X}] \rangle_{(\beta, \vec\kappa,\vec\mu)} = \frac{1}{Z(\beta,\vec\kappa,\vec\mu) } 
\int\! {\cal O} [X] \, w(X; \beta, \vec\kappa, \vec\mu) \, dX .
\label{eq:expop}
\end{eqnarray}

The coupling parameters $(\beta, \vec\kappa, \vec\mu)$ in the histogram $w$ can be shifted by the reweighting technique as
\begin{eqnarray}
\frac{w(X; \beta, \vec\kappa, \vec\mu)}{w(X; \beta_0, \vec\kappa_0, \vec0) }
&=& \left\langle e^{6 (\beta - \beta_0) N_{\rm site} \hat{P}} \,
\prod_f \frac{ \det M(\kappa_f, \mu_f)}{\det M(\kappa_{0f}, 0)} 
\right\rangle_{\! X;(\beta_0, \vec\kappa_0, \vec0)} 
\equiv R(X; \beta, \vec\kappa, \vec\mu; \beta_0, \vec\kappa_0),
\label{eq:R}
\end{eqnarray}
where $\langle \,\cdots\, \rangle_{X;(\beta_0, \vec\kappa_0, 0)}$ is the expectation value measured at $(\beta_0, \vec\kappa_0, \vec\mu_0=0)$ with fixed $X$, 
\begin{eqnarray}
\langle \,\cdots\, \rangle_{X;(\beta_0,\vec\kappa_0,\vec0)}
\equiv \frac{ \langle \,\cdots\, \prod_i \delta(X_i - \hat{X}_i) \rangle_{(\beta_0, \vec\kappa_0, \vec0)} 
}{ \langle \prod_i \delta(X_i - \hat{X}_i) \rangle_{(\beta_0, \vec\kappa_0, \vec0)} } .
\end{eqnarray}
The choice $\vec\mu_0=0$ enables us to carry out  a Monte Carlo simulation of the system.
Note that, when we choose the plaquette $P$ as one of $X_i$'s, the $\beta$-dependence of the reweighing factor $R$ can be simply factored out as 
$e^{6 (\beta - \beta_0) N_{\rm site} P}$. 

For convenience, we define the effective potential as
\begin{equation}
V_{\rm eff}(X; \beta, \vec\kappa, \vec\mu) = -\ln w(X; \beta, \vec\kappa, \vec\mu) .
\end{equation}
The shift of $(\beta, \vec\kappa, \vec\mu)$ is given by
\begin{eqnarray}
V_{\rm eff}(X; \beta, \vec\kappa, \vec\mu) 
= V_{\rm eff}(X; \beta_0, \vec\kappa_0, \vec0) 
- \ln R(X; \beta, \vec\kappa, \vec\mu; \beta_0, \vec\kappa_0).
\label{eq:logR}
\end{eqnarray}
Choosing $P$ as one of $X_i$'s, the $\beta$-dependence of the effective potential is explicitly given by 
\begin{eqnarray}
V_{\rm eff}(P,X; \beta, \vec\kappa, \vec\mu) 
= V_{\rm eff}(P,X; \beta_0, \vec\kappa, \vec\mu) 
-6 N_{\rm site} (\beta-\beta_0)\, P .
\label{eq:chbeta}
\end{eqnarray}
Here, to make clear that $P$ is chosen as one of $X_i$'s, we explicitly write the argument $P$ and redefine $X$ as remaining $X_i$'s.
As we will discuss later, the $\vec\kappa$ and $\vec\mu$ dependences can also be handled easily in the heavy quark region, by choosing the Polyakov line as one of $X_i$'s. 

\begin{figure}[t] 
\vspace{-4mm}
\centerline{
\includegraphics[width=75mm]{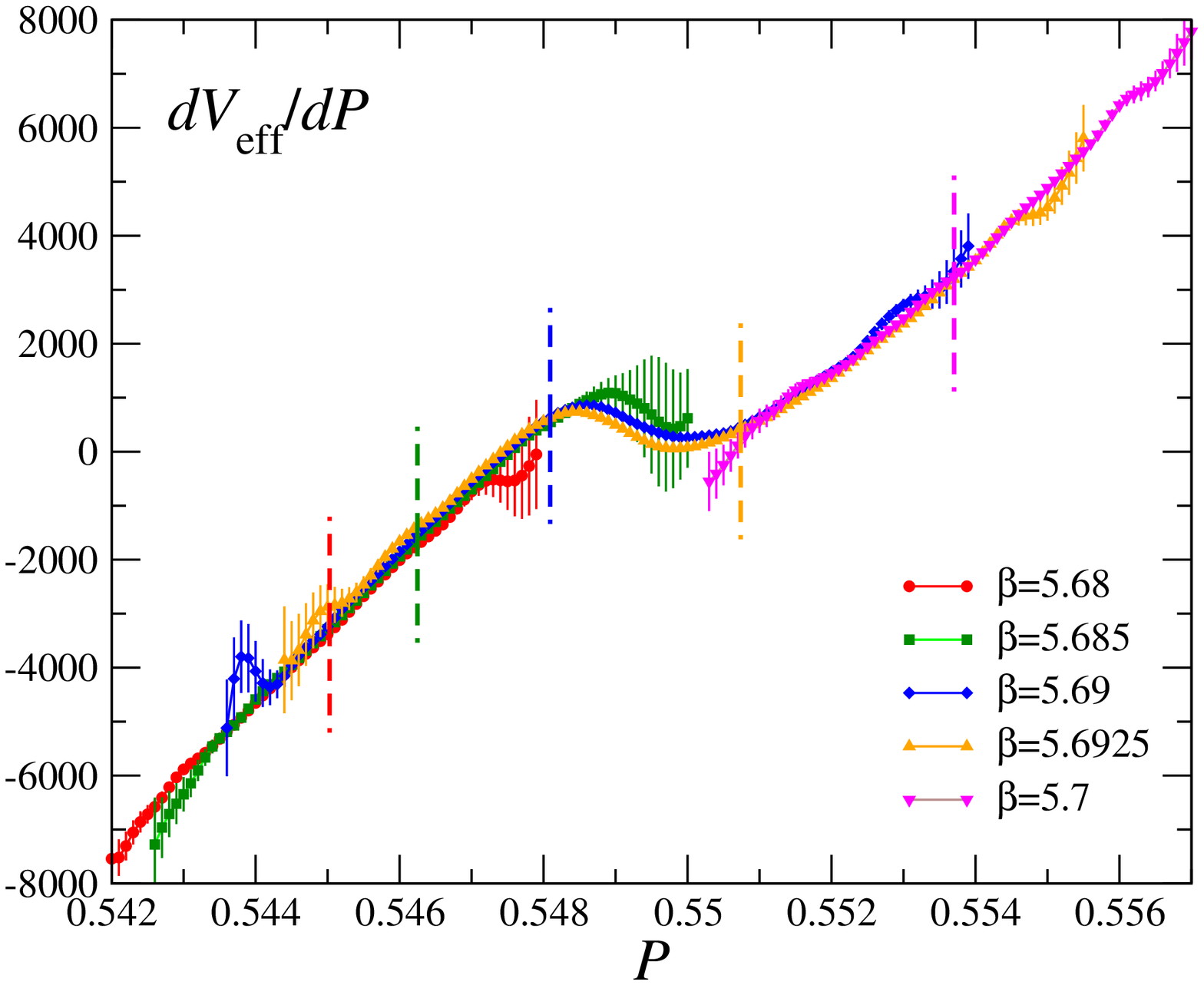} 
\includegraphics[width=75mm]{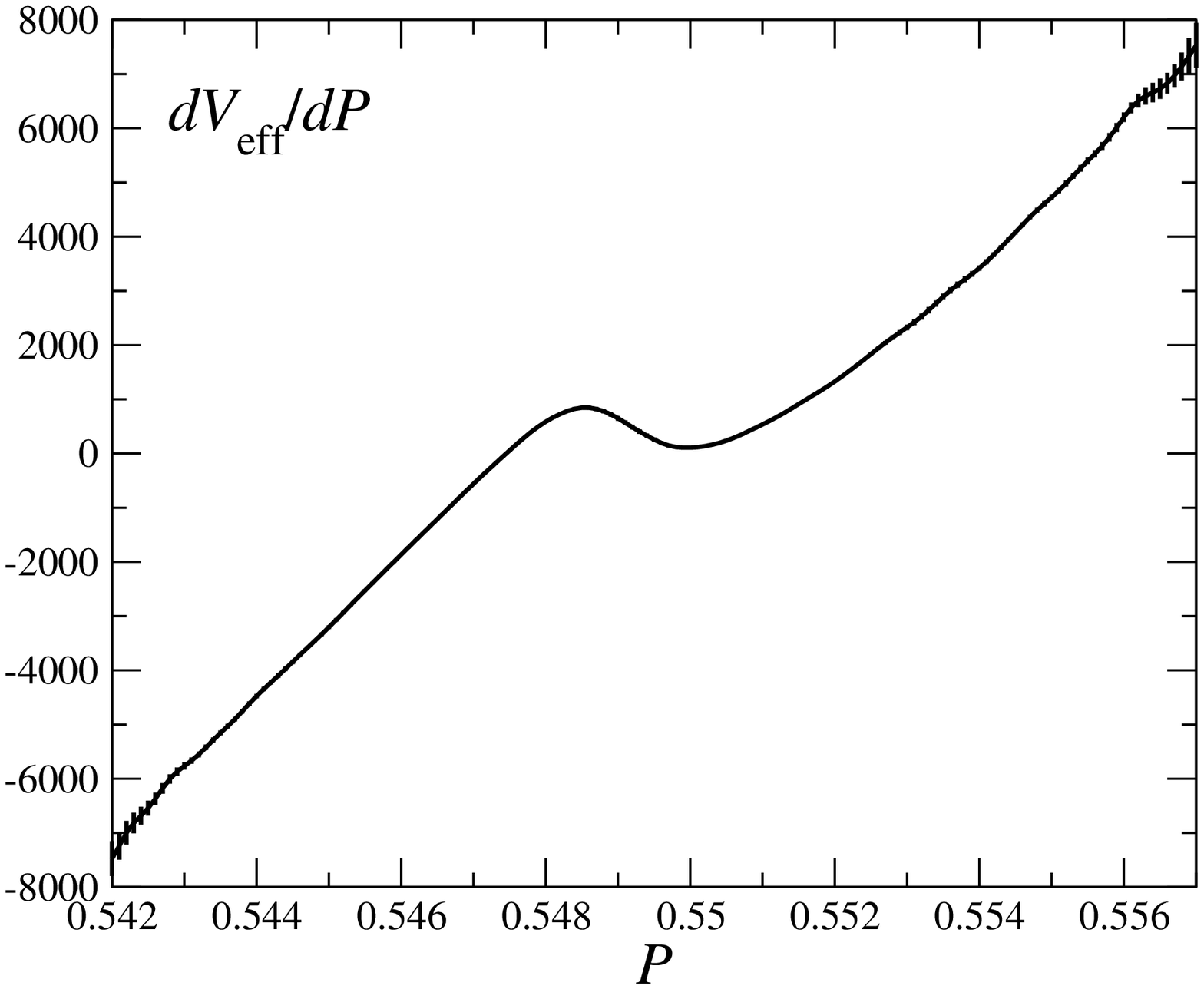} 
}
\caption{Left: $d V_{\rm eff} / dP$ in finite-temperature SU(3) Yang-Mills theory on a $24^3\times4$ lattice \cite{PRD84_054502}. 
Simulations are performed at five values of $\beta$ around the deconfining transition point $\beta_c \simeq 5.692$, and the results for $d V_{\rm eff} / dP$ are shifted to $\beta=5.69$ using the reweighing formula Eq.~(\ref{eq:chbeta}).
The vertical dashed lines represent the expectation values of $P$ for each $\beta$. 
Right: $d V_{\rm eff} / dP$ by the multi-point histogram method.}
\label{fig:dvdp}
\end{figure}

In the left panel of Fig.~\ref{fig:dvdp}, 
$dV_{\rm eff}(P;\beta)/dP$ in finite-temperature SU(3) Yang-Mills theory (heavy quark limit of QCD) is shown for several $\beta$ points around the deconfining transition point,
i.e., $\beta=5.68$, 5.685, 5.69, 5.6925, and 5.70, on an $N_t=4$ lattice \cite{PRD84_054502} as a function of $P$. 
At each $\beta$, we accumulate $100,000$ -- $670,000$ configurations, 
and the total number of configurations is 1,800,000. 
We evaluate $V_{\rm eff}(P)$ using a Gaussian approximation for the delta function in Eq.~(\ref{eq:dist}), 
$\delta(P) \approx \exp [ -(P/\Delta)^2 ] / (\Delta \sqrt{\pi})$ with $\Delta =0.000283$, 
and calculate the derivative numerically fitting the data of $V_{\rm eff}(P)$ in the range between $P-\epsilon/2$ and $P+\epsilon/2$ by a linear function with $\epsilon = 0.0004$.
Since the plaquette is distributed in the range $0.543 \simle P \simle 0.556$ almost uniformly with these configurations, the number of configurations in a range from $P-\Delta/2$ to $P+\Delta/2$ is about 40,000.

The errors are estimated by a jackknife method. 
According to Eq.~(\ref{eq:chbeta}), $dV_{\rm eff}(P;\beta)/dP$ at different $\beta$ should coincide with each other by a constant shift $6 N_{\rm site} (\beta-\beta_0)$. 
The left panel of Fig.~\ref{fig:dvdp} shows that the data at different $\beta$ actually agree well with each other by this shift.

The overall figure S shape of $dV_{\rm eff}(P;\beta)/dP$ corresponds to the fact that the deconfining transition of SU(3) Yang-Mills theory is of first order.
The vertical dashed lines represent the locations of $\langle P \rangle_{(\beta)}$ for each $\beta$.
We note that the data at each $\beta$ can suffer from large errors when $P$ deviates largely from $\langle P \rangle_{(\beta)}$.
In principle, the reweighting formulas in Eqs.~(\ref{eq:R}) and (\ref{eq:logR}) should enable us to predict the shape of $w$ and $V_{\rm eff}$ at different simulation points. 
In practice, however, because a statistically reliable data of $w$ and $V_{\rm eff}$ are available only at $X \approx \langle X \rangle_{(\beta_0,\vec\kappa_0,\vec0)}$ of the simulation point, when $\langle X \rangle_{(\beta,\vec\kappa,\vec\mu)}$ at the target point $(\beta,\vec\kappa,\vec\mu)$ shifts a lot from $\langle X \rangle_{(\beta_0,\vec\kappa_0,\vec0)}$, it is not easy to obtain a reliable prediction about the nature of the vacuum at $(\beta,\vec\kappa,\vec\mu)$.
This is the overlap problem.
For example, in the left panel of Fig.~\ref{fig:dvdp}, it is difficult to conclude about the transition point from the data at $\beta=5.685$ (green) alone.
The overlap problem becomes severe around a first order transition point on large lattices.

\subsection{Multi-point histogram method}
\label{sec:multipoint}

To overcome the overlap problem and thus to obtain $w$ and $V_{\rm eff}$ that are reliable in a wide range of $X$, we make use of the reweighing formulas to combine data obtained at different simulation points \cite{FS89}. 
In this subsection, we consider the case to combine data at different values of $\beta$ with fixed $\vec\kappa$ and $\vec\mu$ and suppress the arguments $\vec\kappa$ and $\vec\mu$ for simplicity of the notations. 
Extension to the general case is straightforward.

We combine a set of $N_{\rm sp}$ simulations performed at $\beta_i$ with the number of configurations $N_i$ where $i=1, \cdots , N_{\rm sp}$.
Using Eq.~(\ref{eq:chbeta}), the probability distribution function at $\beta_i$ is related to that at $\beta$ as  
\[
Z^{-1}(\beta_i) \,w(P,X; \beta_i) = Z^{-1}(\beta_i) \, e^{6N_{\rm site} (\beta_i-\beta) P}  \,w(P, X; \beta) .
\]
Summing up these probability distribution functions with the weight $N_i$, 
\begin{eqnarray}
\sum_{i=1}^{N_{\rm sp}} N_i \, Z^{-1}(\beta_i) \, w(P,X; \beta_i) 
= e^{-6N_{\rm site} \beta P} 
\sum_{i=1}^{N_{\rm sp}} N_i \, Z^{-1}(\beta_i) \, e^{6N_{\rm site}  \beta_i P}  \, w(P, X; \beta), 
\label{eq:sum1}
\end{eqnarray}
we obtain 
\begin{eqnarray}
w(P, X; \beta)= G(P;\beta,\vec\beta) \,
\sum_{i=1}^{N_{\rm sp}} N_i \, Z^{-1}(\beta_i) \, w(P,X; \beta_i) 
\end{eqnarray}
with the simulation points $\vec\beta=(\beta_1,\cdots,\beta_{N_{\rm sp}})$ and 
\begin{eqnarray}
G(P;\beta,\vec\beta)=\frac{ e^{6N_{\rm site} \beta P}}{
\sum_{i=1}^{N_{\rm sp}} N_i \, e^{6N_{\rm site} \beta_i P} Z^{-1}(\beta_i)} .
\end{eqnarray}
Note that the left-hand side of Eq.~(\ref{eq:sum1}) gives a naive histogram using all the configurations disregarding the difference in the simulation parameter.
The histogram $w(P,X;\beta)$ at $\beta$ is given by multiplying $G(P;\beta,\vec\beta)$ to this naive histogram.

The partition function is given by
\begin{eqnarray}
Z(\beta)= \sum_{i=1}^{N_{\rm sp}} N_i \int G(P;\beta,\vec\beta) \, Z^{-1}(\beta_i) \, w(P,X; \beta_i) \, dP \,dX 
=\sum_{i=1}^{N_{\rm sp}} N_i \left\langle G(\hat{P};\beta,\vec\beta) \right\rangle_{\!(\beta_i)}.
\end{eqnarray}
The right-hand side is just the naive sum of $G(\hat{P};\beta,\vec\beta)$ observed on all the configurations.
The partition function at $\beta_i$ can be determined, up to an overall factor, by the consistency relations, 
\begin{eqnarray}
Z(\beta_i) 
=\sum_{k=1}^{N_{\rm sp}} N_k \left\langle G(\hat{P};\beta_i,\vec\beta) \right\rangle_{\! (\beta_k)}
=\sum_{k=1}^{N_{\rm sp}} N_k \left\langle 
\frac{e^{6N_{\rm site} \beta_i P}}{
\sum_{j=1}^{N_{\rm sp}} N_j e^{6N_{\rm site} \beta_j P} Z^{-1}(\beta_j)} \right\rangle_{\! (\beta_k)}
\end{eqnarray}
for $i=1,\cdots,N_{\rm sp}$. 
Denoting $f_i=-\ln Z(\beta_i)$, these equations can be rewritten by
\begin{eqnarray}
1 = \sum_{k=1}^{N_{\rm sp}} N_k \left\langle
\frac{1}{ \sum_{j=1}^{N_{\rm sp}} N_j \exp[ 6N_{\rm site} (\beta_j-\beta_i) P 
- f_i +f_j]} \right\rangle_{(\beta_k)},
\hspace{5mm}
i=1,\cdots,N_{\rm sp}.
\label{eq:consis}
\end{eqnarray}
Starting from appropriate initial values of $f_i$, we solve these equations numerically by an iterative method. 
Note that, in these calculations, one of the $f_i$'s must be fixed to remove the ambiguity corresponding to the undetermined overall factor.

Now, the expectation value of an operator ${\cal O}[\hat{P},\hat{X}]$ at $\beta$ can be evaluated as
\begin{eqnarray}
\langle {\cal O} [\hat{P},\hat{X}] \rangle_{(\beta)} 
= \frac{1}{Z(\beta)} \sum_{i=1}^{N_{\rm sp}} N_i \left\langle {\cal O} [\hat{P},\hat{X}] \, G(\hat{P};\beta,\vec\beta) \right\rangle_{\!(\beta_i)}.
\label{eq:multibeta}
\end{eqnarray}
Again, 
$\sum_{i=1}^{N_{\rm sp}} N_i \left\langle {\cal O} G \right\rangle_{(\beta_i)}$ 
in the right-hand side is just the naive sum of ${\cal O} G$ over all the configurations disregarding the difference in the simulation point.

\begin{figure}[t] 
\vspace{-4mm}
\centerline{
\includegraphics[width=120mm]{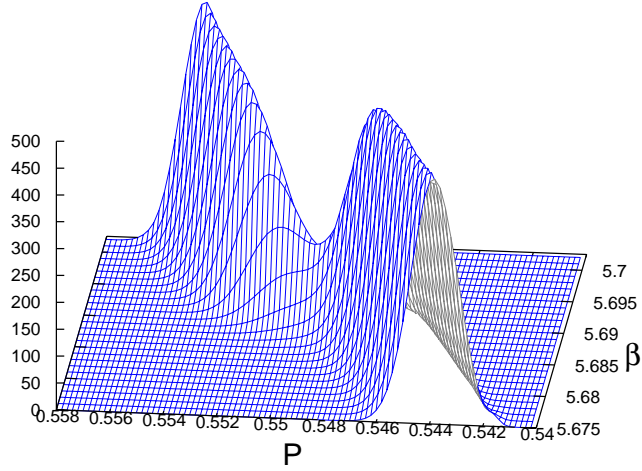} 
}
\vspace{-8mm}
\caption{$\beta$ dependence of the plaquette distribution function at $\kappa=0$.}
\label{fig:pqdist}
\end{figure}

Results of the multi-point histogram method for the plaquette distribution function $Z^{-1}(\beta)\, w(P;\beta)$ in SU(3) Yang-Mills theory is shown in Fig.~\ref{fig:pqdist} as a function of $P$ and $\beta$, using the configurations at five simulation points presented in the left panel of Fig.~\ref{fig:dvdp}.
We see double-peak distribution at $\beta \sim 5.69$--5.6925.
Corresponding $dV_{\rm eff}(P;\beta)/dP$ at $\beta=5.69$ is shown in the right panel of Fig.~\ref{fig:dvdp}.
We obtain a smooth effective potential with small statistical errors in a wide range of $P$, automatically suppressing statistically poor data points by the multi-point histogram method.

\section{QCD in the heavy quark region}
\label{sec:heavy}

In this paper, we employ the standard plaquette gauge action, Eq.~(\ref{eq:Sg}) with Eq.~(\ref{eq:SgP}), and unimproved Wilson quark action, Eq.~(\ref{eq:Sq}), with
\begin{eqnarray}
M_{xy} (\kappa_f,\mu_f) &=& \delta_{xy} 
-\kappa_f \left\{ \sum_{\mu=1}^3\left[ (1-\gamma_{\mu})\,U_{x,\mu}\,\delta_{y,x+\hat{\mu}} + (1+\gamma_{\mu})\,U_{y,\mu}^{\dagger}\,\delta_{y,x-\hat{\mu}} \right]  \right. 
\nonumber  \\
        & & \left. + e^{\mu_fa}(1-\gamma_{4})\,U_{x,4}\,\delta_{y,x+\hat{4}}+e^{-\mu_fa}(1+\gamma_{4})\,U_{y,4}^{\dagger}\,\delta_{y,x-\hat{4}} \right\} 
.
\end{eqnarray} 
Note that this $M_{xy}$ does not depend on $\beta$.

To investigate the quark mass dependence of the effective potential in the heavy quark region, we evaluate the quark determinant by a Taylor expansion with respect to the set of hopping parameters $\vec\kappa$ in the vicinity of the heavy quark limit $\vec\kappa=0$.
For each flavor, we have 
\begin{eqnarray}
\ln \left[ \frac{\det M(\kappa,\mu)}{\det M(0,0)} \right]
\;=\; \sum_{n=1}^{\infty} \frac{1}{n!} 
\left[ \frac{\partial^{n} \ln \det M}{\partial \kappa^{n}} 
\right]_{\kappa=0} \kappa^{n} 
\;=\; \sum_{n=1}^{\infty} \frac{{\cal D}_{n}}{n!}  \, \kappa^{n} , 
\label{eq:tayexp}
\end{eqnarray}
with
\begin{eqnarray}
{\cal D}_n &\equiv& \left[ \frac{\partial^n \ln \det M}{\partial \kappa^n} \right]_{\kappa=0}
\nonumber\\
&=& (-1)^{n+1} (n-1)! \ {\rm tr} 
\left[ \left( M^{-1} \, \frac{\partial M}{\partial \kappa} \right)^n \right]_{\kappa=0}
=(-1)^{n+1} (n-1)! \ {\rm tr} 
\left[ \left( \frac{\partial M}{\partial \kappa} \right)^n \right], 
\label{eq:derkappa}
\end{eqnarray}
where $(\partial M/\partial \kappa)_{xy}$ is the gauge connection between the sites $x$ and $y$.
Therefore, the nonvanishing contributions to ${\cal D}_{n}$ are given by Wilson loops and Polyakov lines. 
Because QCD at $\vec\kappa=0$ is just the SU(3) Yang-Mills theory, we can easily perform simulations at $\vec\kappa=0$.

Because of the antiperiodic boundary condition and gamma matrices in the hopping terms, the leading order contributions to the Taylor expansion are given by 
\begin{equation}
\ln \left[ \frac{\det M(\kappa, \mu)}{\det M(0,0)} \right]
  = 288N_{\rm site} \kappa^4 \hat{P} +3\times2^{N_t+2}N_s^3\kappa^{N_t} 
\left\{ \cosh \left(\frac{\mu}{T}\right) \hat\Omega_{\rm R}
+i\sinh \left( \frac{\mu}{T} \right) \hat\Omega_{\rm I}\right\} + \cdots ,
\label{eq:detM}
\end{equation}
where $\hat\Omega_{\rm R}$ and $\hat\Omega_{\rm I}$ are the real and imaginary parts of the Polyakov line, 
\begin{equation}
\hat\Omega = \frac{1}{N_s^3}
\displaystyle \sum_{\mathbf{n}} \frac{1}{3} {\rm tr} \left[ 
U_{\mathbf{n},4} U_{\mathbf{n}+\hat{4},4} U_{\mathbf{n}+2\hat{4},4} 
\cdots U_{\mathbf{n}+(N_t -1)\hat{4},4} \right] ,
\label{eq:ploop}
\end{equation}
respectively.
Collecting the contributions from all flavors, we have 
\begin{eqnarray}
\ln \left[ \prod_{f=1}^{N_{\rm f}} \frac{\det M(\kappa_f, \mu_f)}{\det M(0,0)} \right] 
&=& 288N_{\rm site} \sum_{f=1}^{N_{\rm f}} \kappa_f^4 \hat{P} 
\nonumber \\ 
&& \hspace{-35mm} 
+ 3\times2^{N_t+2}N_s^3 \left\{ 
\sum_{f=1}^{N_{\rm f}} \kappa_f^{N_t} \cosh \left(\frac{\mu_f}{T}\right) \hat\Omega_{\rm R}
+i \sum_{f=1}^{N_{\rm f}} \kappa_f^{N_t} \sinh \left( \frac{\mu_f}{T} \right) 
\hat\Omega_{\rm I}\right\} + \cdots .
\label{eq:proddetM}
\end{eqnarray}
The first term that is proportional to $\hat{P}$ can be absorbed into the gauge action by a shift $\beta \rightarrow \beta^*$ with 
\begin{eqnarray}
\beta^* = \beta + 48 \sum_{f=1}^{N_{\rm f}} \kappa_f^4 .
\label{eq:betastar}
\end{eqnarray}
The third term that is proportional to $\hat\Omega_{\rm I}$ leads to the complex phase factor $e^{ i\hat\theta}$,
where
\begin{equation}
\hat\theta=3\times 2^{N_t+2} N_s^3 
\, q \,  
\hat\Omega_{\rm I} 
\label{eq:thetaOmegaI}
\end{equation}
with
\begin{eqnarray}
q = \sum_{f=1}^{N_{\rm f}} \kappa_f^{N_t} \sinh \left( \frac{\mu_f}{T} \right) .
\label{eq:q}
\end{eqnarray}
From these expressions, it is natural to take the Polyakov line as an argument of effective potential.

\section{Polyakov line effective potential at zero density}
\label{sec:pldist0}

\begin{figure}[t] 
\vspace{-4mm}
\centerline{
\includegraphics[width=90mm]{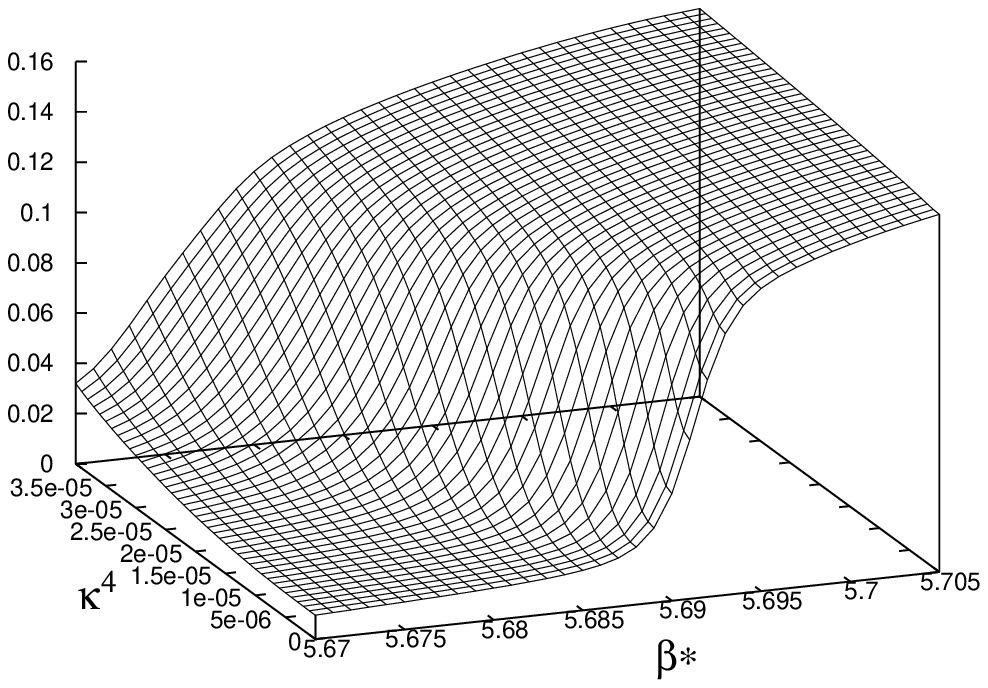} 
\hspace{-15mm}
\includegraphics[width=90mm]{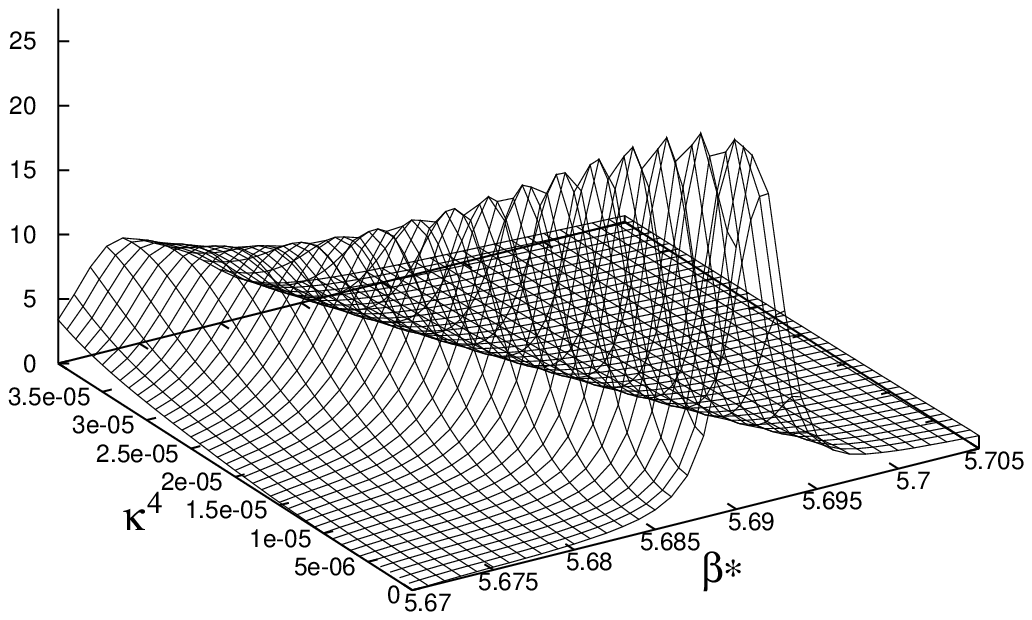} 
}
\vspace{-4mm}
\caption{Polyakov line expectation value $\langle \hat\Omega \rangle$  (left) and its susceptibility $\chi_{\Omega}$ (right) as functions of $\kappa^4$ and $\beta^*$ at $\vec\mu=0$ obtained on an $N_t=4$ lattice in the case of $N_{\rm f}=2$ QCD.}
\label{fig:plsus}
\end{figure}

\begin{figure}[t] 
\centerline{
\includegraphics[width=75mm]{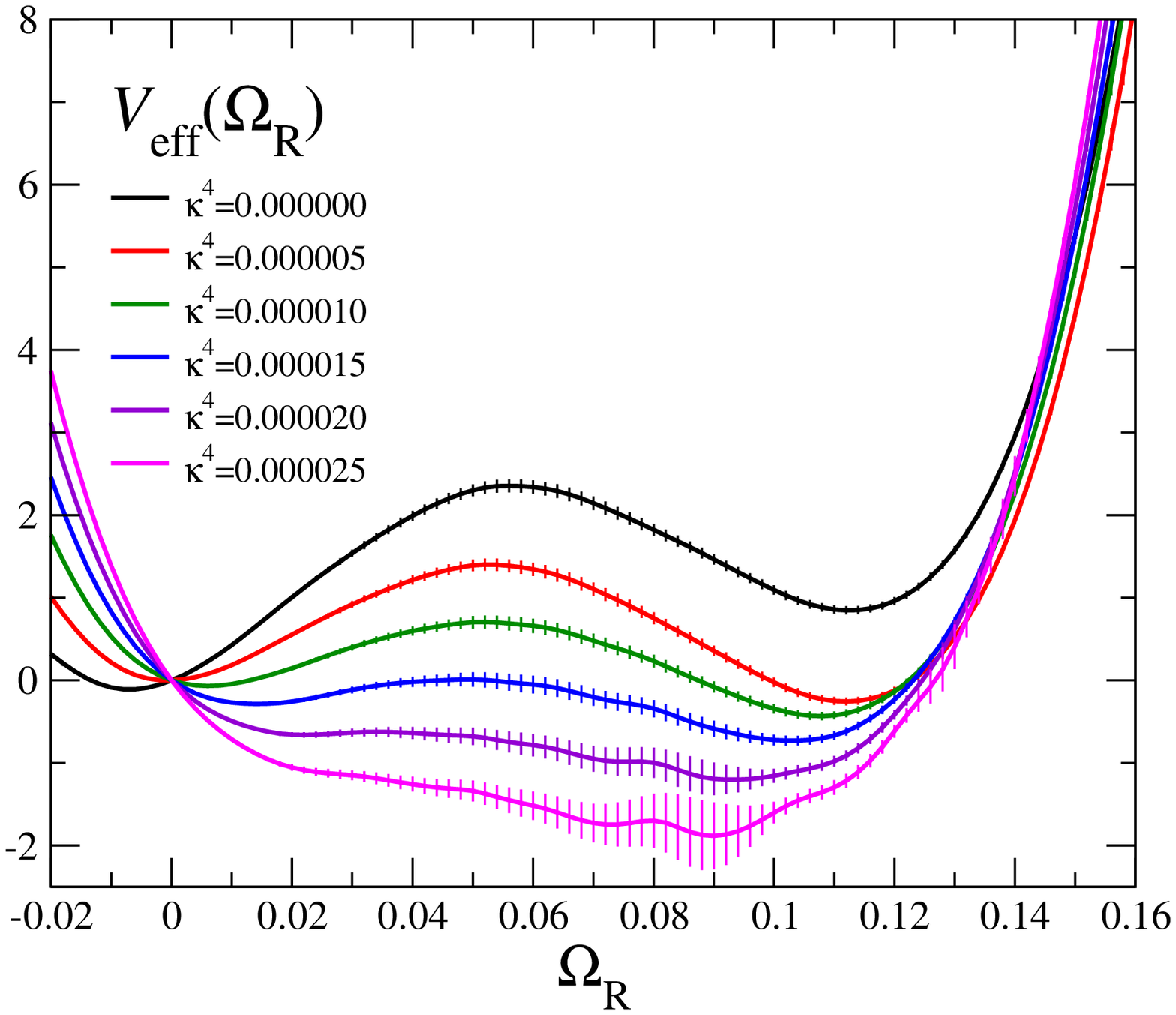}
\includegraphics[width=75mm]{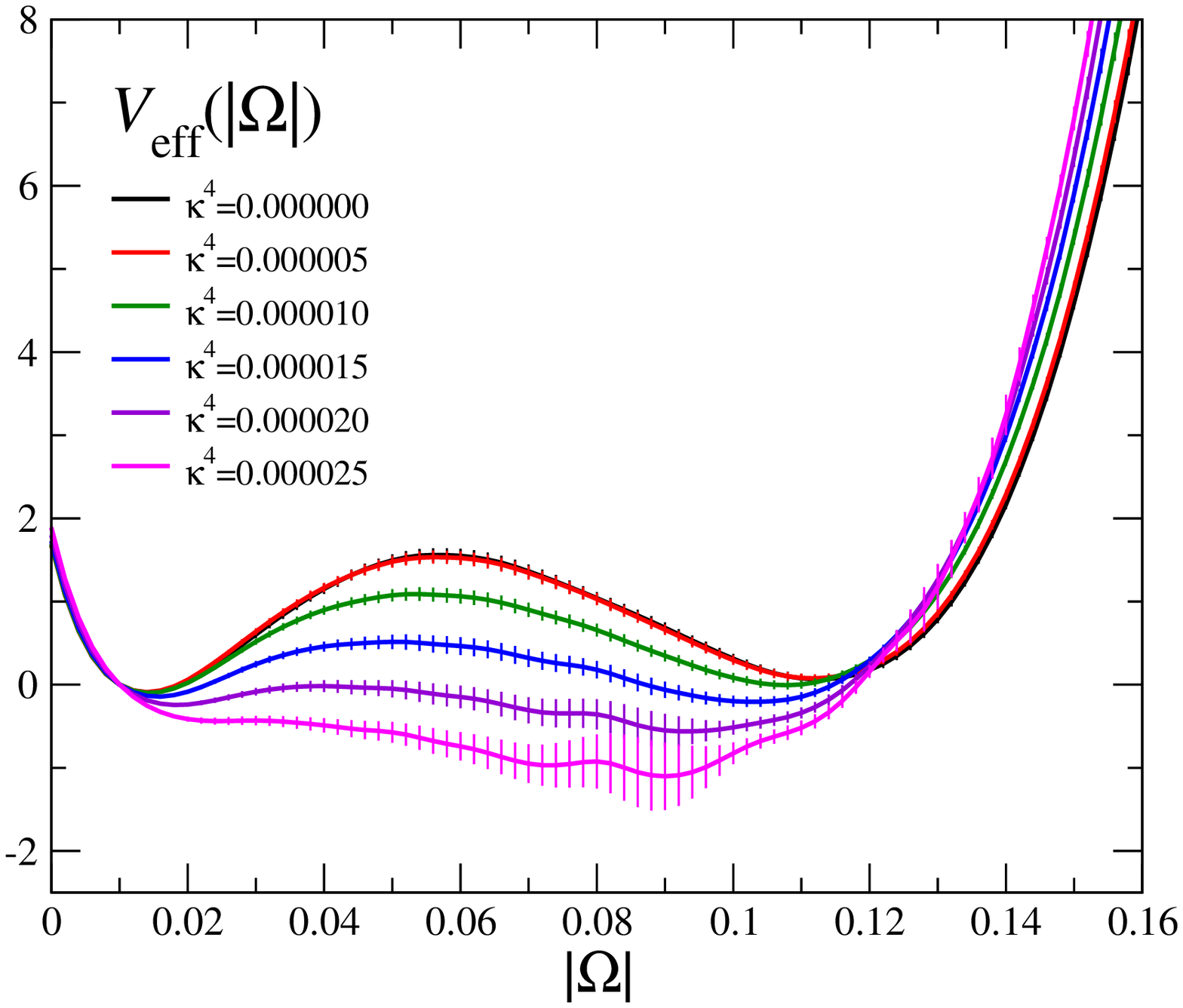}
}
\caption{Distribution function of the real part of the Polyakov line (left) and 
the absolute value of the Polyakov line (right) at the transition point in $N_{\rm f}=2$ QCD at $\mu=0$.}
\label{fig:oabhist}
\end{figure}

\begin{figure}[t] 
\centerline{
\includegraphics[width=75mm]{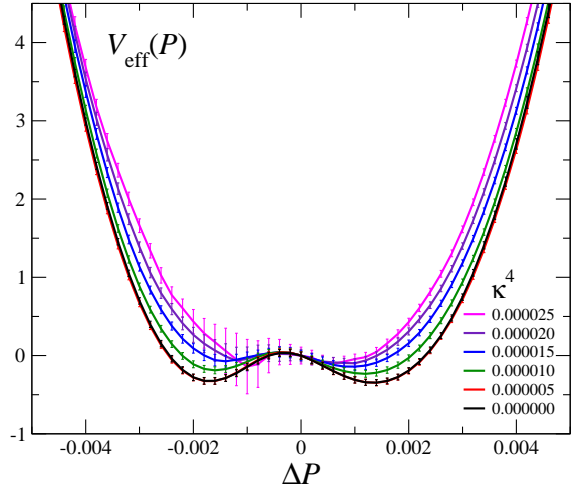}
}
\caption{Plaquette effective potential in $N_{\rm f}=2$ QCD at $\mu=0$. 
The horizontal axis is for $\Delta P = P - \langle \hat P \rangle$.
$\beta$ is adjusted to the peak position of the plaquette susceptibility at each $\kappa$.}
\label{fig:phist}
\end{figure}

\begin{figure}[tb]
\vspace{-9mm}
\centerline{
\includegraphics[width=59mm]{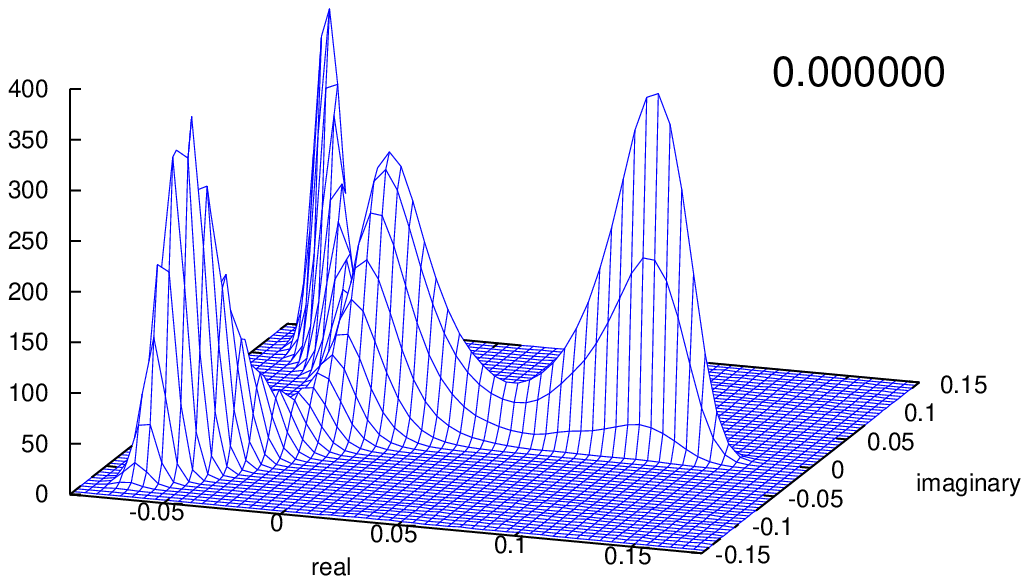}
\hspace{-11mm}
\includegraphics[width=59mm]{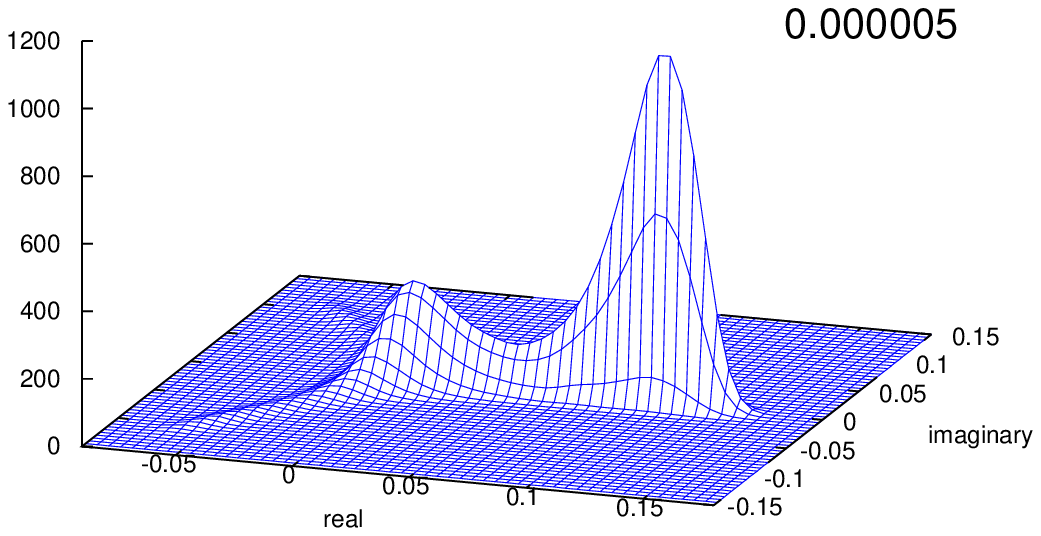}
\hspace{-11mm}
\includegraphics[width=59mm]{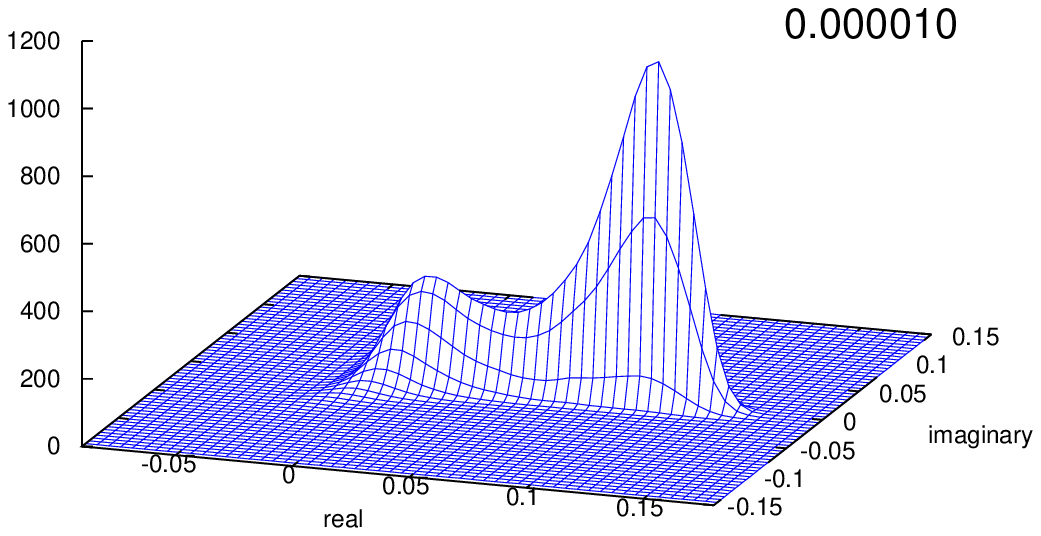}
}
\vspace{-12mm}
\centerline{
\includegraphics[width=59mm]{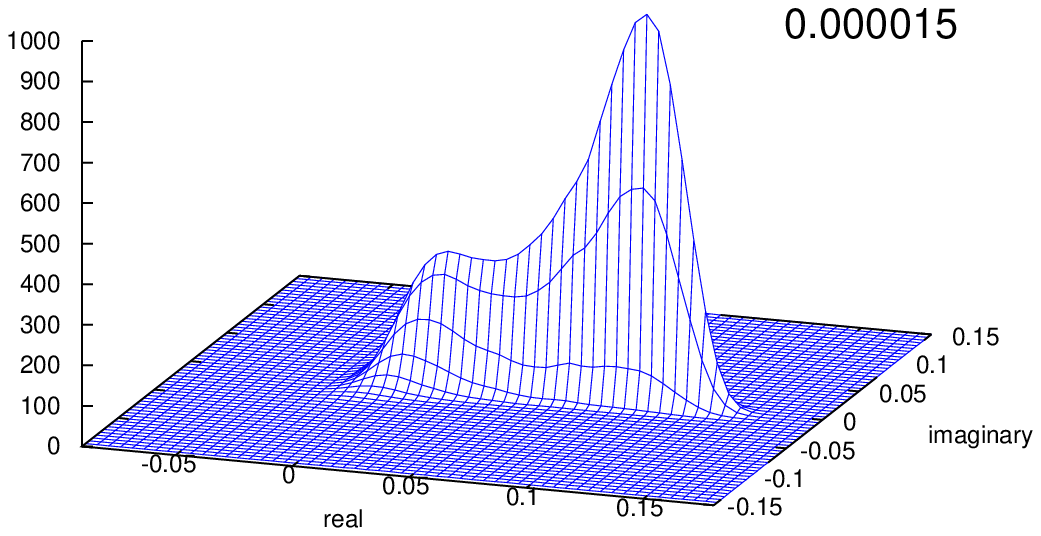}
\hspace{-11mm}
\includegraphics[width=59mm]{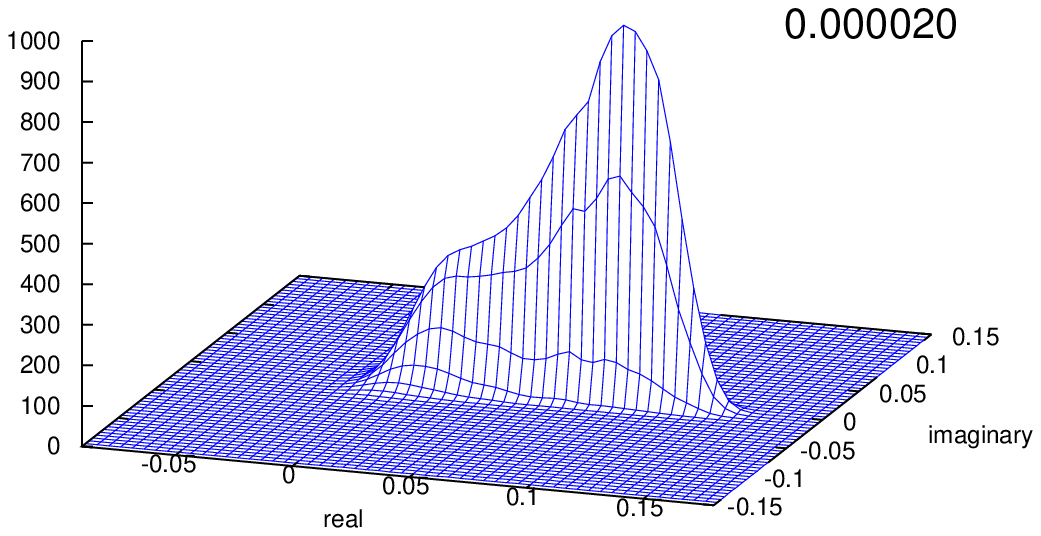}
\hspace{-11mm}
\includegraphics[width=59mm]{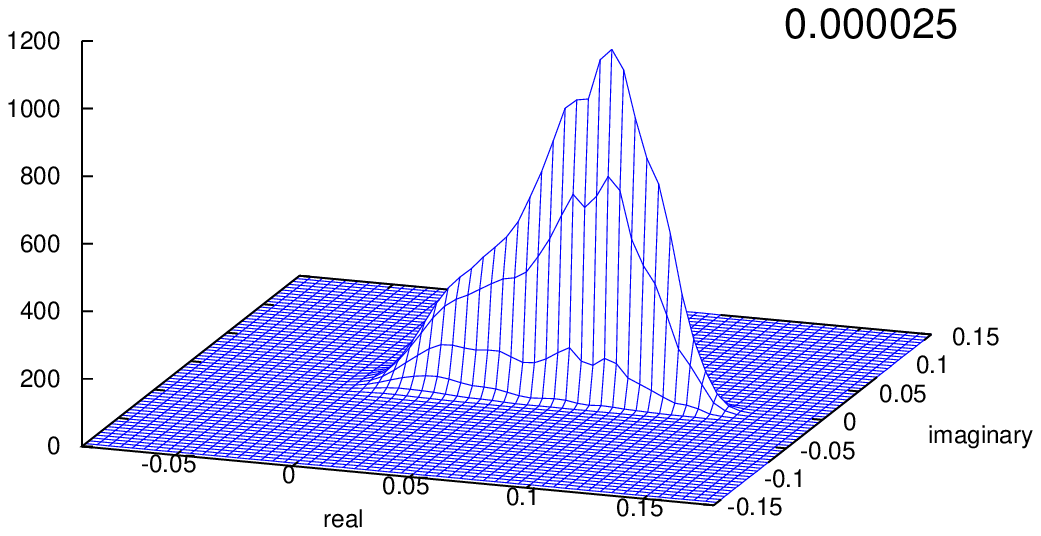}
}
\vspace{-2mm}
\caption{Polyakov line histogram in heavy quark QCD at $\vec\mu=0$ in the case of $N_{\rm f}=2$ QCD.
At each $\kappa$, $\beta$ is adjusted to the transition point determined by $\chi_\Omega$.
The value of $\kappa^4$ is shown in the upper right corner of each plot.
}
\label{fig:compol}
\end{figure}

We expect that the Polyakov line $\hat\Omega$ is sensitive to the transition in the heavy quark region because it is an order parameter of the confinement-deconfinement transition in SU(3) Yang-Mills theory, the heavy quark limit of QCD.
Combining the configurations at five simulation points \cite{PRD84_054502} used in the left panel of Fig.~\ref{fig:dvdp} by the multi-point reweighing formulas, 
we compute the Polyakov line histogram in the heavy quark region. 
To reduce statistical fluctuations, we average $\Omega$ over the Z(3) rotations in the original histogram at $\vec\kappa=0$ (SU(3) Yang-Mills theory).

We first study the case $\vec\mu=0$ in this section.
In Fig.~\ref{fig:plsus} we show the Polyakov line expectation value $\langle \hat\Omega \rangle$ and its susceptibility $\chi_{\Omega} =N_{s}^3 \langle (\hat\Omega - \langle \hat\Omega \rangle)^2 \rangle$ in the case of $N_{\rm f}=2$ QCD (degenerate two-flavor QCD).
$\langle \hat\Omega \rangle$ is real, but $\chi_{\Omega}$ includes the fluctuations in the imaginary part too.
We define the transition point as the peak position of $\chi_{\Omega}$.
Owing to the multi-point reweighting method, these quantities can be calculated in a wide range of $\beta$ and $\kappa$. 

To compute the histogram for the Polyakov line, we approximate $\delta(x) \approx \exp[-(x/\Delta)^2]/(\Delta \sqrt{\pi}) $ as in the case of the plaquette histogram, but now with $\Delta=0.005$ considering the resolution and the statistical error. 
The effective potential for $\Omega_{\rm R}$ is shown in the left panel of Fig.~\ref{fig:oabhist}, where $\beta$ is adjusted to the transition point at each $\kappa$.  
In this plot, we vertically shift $V_{\rm eff}(\Omega_{\rm R})$ by adding a constant at each $\kappa$ such that $V_{\rm eff}(\Omega_{\rm R}\!=\!0)=0$.
We find the critical point where the first order transition turns into a crossover at $\kappa_{\rm cp}^4 \approx 0.00002$ ($\kappa_{\rm cp} \approx 0.0669$) in $N_{\rm f} = 2$ QCD.
In a study of SU(3) Yang-Mills theory, the absolute value of the Polyakov line $|\Omega|$ is also used to detect the transition. 
Using the same approximation for the delta function, we obtain $V_{\rm eff}(|\Omega|)$ shown in the right panel of Fig.~\ref{fig:oabhist}, where $V_{\rm eff}(|\Omega|)$ is vertically shifted such that $V_{\rm eff}(|\Omega|)=0$ at $|\Omega|=0.01$.
We find that $\kappa_{\rm cp}^4 \approx 0.00002$ also in this determination. 

In Ref.~\cite{PRD84_054502}, $\kappa_{\rm cp}=0.0658(3)(^{+4}_{-11})$ with $\beta^*_{\rm cp} = 5.6836(1)(5)$ ($\beta_{\rm cp}=5.6819(1)(5)$) was obtained for $N_{\rm f} = 2$ QCD, combining the results of three different analyses on the plaquette effective potential, where the first bracket is for the statistical error and the second bracket is for the systematic error estimated by the analysis method dependence. 
Our results of $\kappa_{\rm cp}$ from the Polyakov line effective potential are consistent with this estimation. 
Figure~\ref{fig:phist} is an update of Fig.~4 in Ref.~\cite{PRD84_054502} by using the multi-point histogram method.
In this plot, $\beta$ is adjusted to the peak position of the plaquette susceptibility at each $\kappa$, and $V_{\rm eff}(P)$ is vertically shifted at each $\kappa$ such that $V_{\rm eff}(P\!=\!\langle \hat P \rangle )=0$.

Finally, we show the histogram of the Polyakov line in the complex $\Omega=(\Omega_{\rm R}, \Omega_{\rm I})$ plane in Fig.~\ref{fig:compol}.
At each $\kappa$, $\beta$ is adjusted to the first-order transition point determined by $\chi_\Omega$.
At $\kappa=0$, we find a peak at $\Omega\approx0$ corresponding to the low-temperature confined phase and three peaks corresponding to the high-temperature deconfined phase in which the center Z(3) symmetry is spontaneously broken.
At $\kappa \ne 0$, the Z(3) symmetry is explicitly violated so that the branch on the real axis of the complex $\Omega$ plane is selected.

Before closing this section, let us discuss the effects of higher order terms of the hopping parameter expansion.
At higher orders of the expansion, more complicated loops contribute to the reweighing factor besides the plaquette and Polyakov line.
In Appendix~\ref{sec:range}, we examine the effects of next-to-leading order $\kappa^6$ terms on the evaluation of the critical point at $\mu=0$.
We find that the effects of $\kappa^6$ loops are small around the critical point and the shift of $\kappa_{\rm cp}$ due to the next-to-leading order terms is just about 3\% at $N_t=4$.

\section{Polyakov line effective potential at finite density}
\label{sec:poldis}

We now study the case of finite $\mu$.
As discussed in Sec.~\ref{sec:heavy}, to the leading order of the hopping parameter expansion, finite $\mu$ has two effects: 
(i) the shift of the effective coupling for $\hat\Omega_{\rm R}$ and (ii) the introduction of the complex phase factor.
According to Eqs.~(\ref{eq:R}) and (\ref{eq:proddetM}), the histogram of $\Omega_{\rm R}$ is given by
\begin{eqnarray}
&& \hspace{-7mm} w(\Omega_{\rm R}; \beta, \vec\kappa, \vec\mu) 
\nonumber \\
&=& w(\Omega_{\rm R}; \beta, \vec0, \vec0) 
\times \nonumber\\ &&
\left\langle  e^{288N_{\rm site} \sum_{f=1}^{N_{\rm f}} \kappa_f^4 \hat{P}} 
\exp \left[ 3 N_s^3 2^{N_t+2} \sum_{f=1}^{N_{\rm f}}
\kappa_f^{N_t}  \left\{ \cosh \left( \frac{\mu_f}{T} \right) \hat{\Omega}_{\rm R}
+i\sinh \left( \frac{\mu_f}{T} \right) \hat{\Omega}_{\rm I} \right\} \right]
\right\rangle_{\Omega_{\rm R}; (\beta, \vec0,\vec0)} 
\nonumber \\
&=& w(\Omega_{\rm R}; \beta^*, \vec0, \vec0) 
e^{ 3 N_s^3 2^{N_t+2} \, h \, \Omega_{\rm R} } \times
\left\langle e^{i \hat{\theta}} \right\rangle_{\Omega_{\rm R}; (\beta^*, \vec0,\vec0)} 
\label{eq:plhvdist}
\end{eqnarray}
with 
\begin{eqnarray}
h = \sum_{f=1}^{N_{\rm f}} \kappa_f^{N_t} \cosh \left( \frac{\mu_f}{T} \right) .
\label{eq:h}
\end{eqnarray}
$\beta^*$ and $\hat{\theta}$ are defined by Eqs.~(\ref{eq:betastar}), (\ref{eq:thetaOmegaI}) and (\ref{eq:q}).
We note that the histogram depends on the coupling parameters only through $\beta^*$, $h$ and $q$.

\subsection{Phase-quenched finite density QCD}
\label{sec:isospin}

\begin{figure}[t] 
\centerline{
\includegraphics[width=7.5cm]{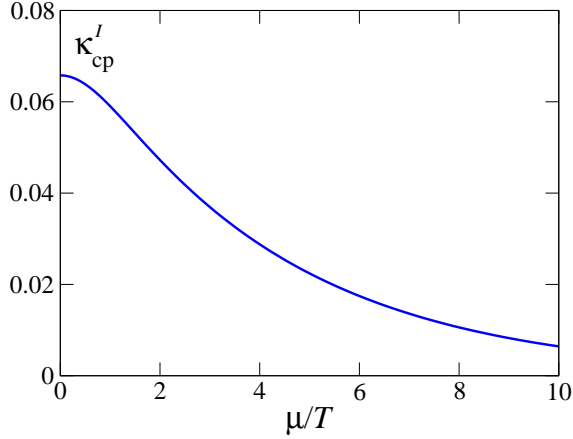} 
}
\caption{Critical point in the phase-quenched $N_{\rm f}=2$ QCD.}
\label{fig:kcp_finitemu}
\end{figure}

It is convenient to first consider the case of phase-quenched finite density QCD, in which the complex phase of the quark determinant is removed.
In $N_{\rm f}=2$ QCD, this corresponds to the case of the isospin chemical potential, $\mu_u=-\mu_d \equiv \mu$. 
Neglecting the complex phase factor in Eq.~(\ref{eq:plhvdist}), we find that $w(\Omega_{\rm R}; \beta, \vec\kappa, \vec\mu)$ in phase-quenched QCD is just the  $w(\Omega_{\rm R}; \beta, \vec\kappa, \vec0)$ with $\kappa_f^{N_t}$ replaced by $\kappa_f^{N_t} \cosh(\mu_f/T)$.
Therefore, e.g., $V_{\rm eff} (\Omega_{\rm R})$ at $\mu=0$ shown in the left panel of Fig.~\ref{fig:oabhist} can be viewed as that at $\mu\ne0$ in the phase-quenched $N_{\rm f}=2$ theory (QCD with isospin chemical potential) by the same replacement of $\kappa$. 

The critical point in the phase-quenched $N_{\rm f}$-flavor QCD is thus given by 
\begin{equation}
\sum_{f=1}^{N_{\rm f}} \left[\kappa_{f;{\rm cp}}^I (\mu_f)\right]^{N_t} \cosh\left(\frac{\mu_f}{T}\right)= 2 \left[\kappa_{\rm cp}(0)\right]^{N_t} , 
\label{eq:isospinf}
\end{equation}
where $\kappa_{\rm cp}(0)$ is the critical point in $N_{\rm f}=2$ QCD at $\mu=0$.
For $N_{\rm f}=2$ QCD with isospin chemical potential $\mu$, the critical point is given by 
\begin{equation}
\kappa_{\rm cp}^I (\mu) = \frac{\kappa_{\rm cp}(0)}{[\cosh(\mu/T)]^{1/N_t} } .
\label{eq:isospin}
\end{equation}
Using the value of $\kappa_{\rm cp}(0)$ determined in the previous section, we plot $\kappa_{\rm cp}^I (\mu)$ in Fig.~\ref{fig:kcp_finitemu}.
Note that, with increasing $\mu$, the critical point approaches $\kappa=0$ where the hopping parameter expansion becomes exact.

\subsection{Complex phase factor by the cumulant expansion method}
\label{sec:phase}
\begin{figure}[t] 
\centerline{
\includegraphics[width=8cm]{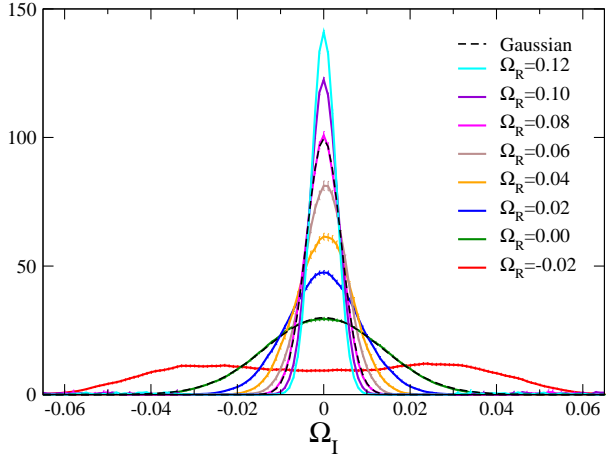} 
}
\caption{The distribution of $\Omega_{\rm I}$ for fixed $\Omega_{\rm R}$ at $\vec\kappa=0$ and $\beta$ at the transition point.}
\label{fig:OmegaI}
\end{figure}

\begin{figure}[t] 
\centerline{
\includegraphics[width=75mm]{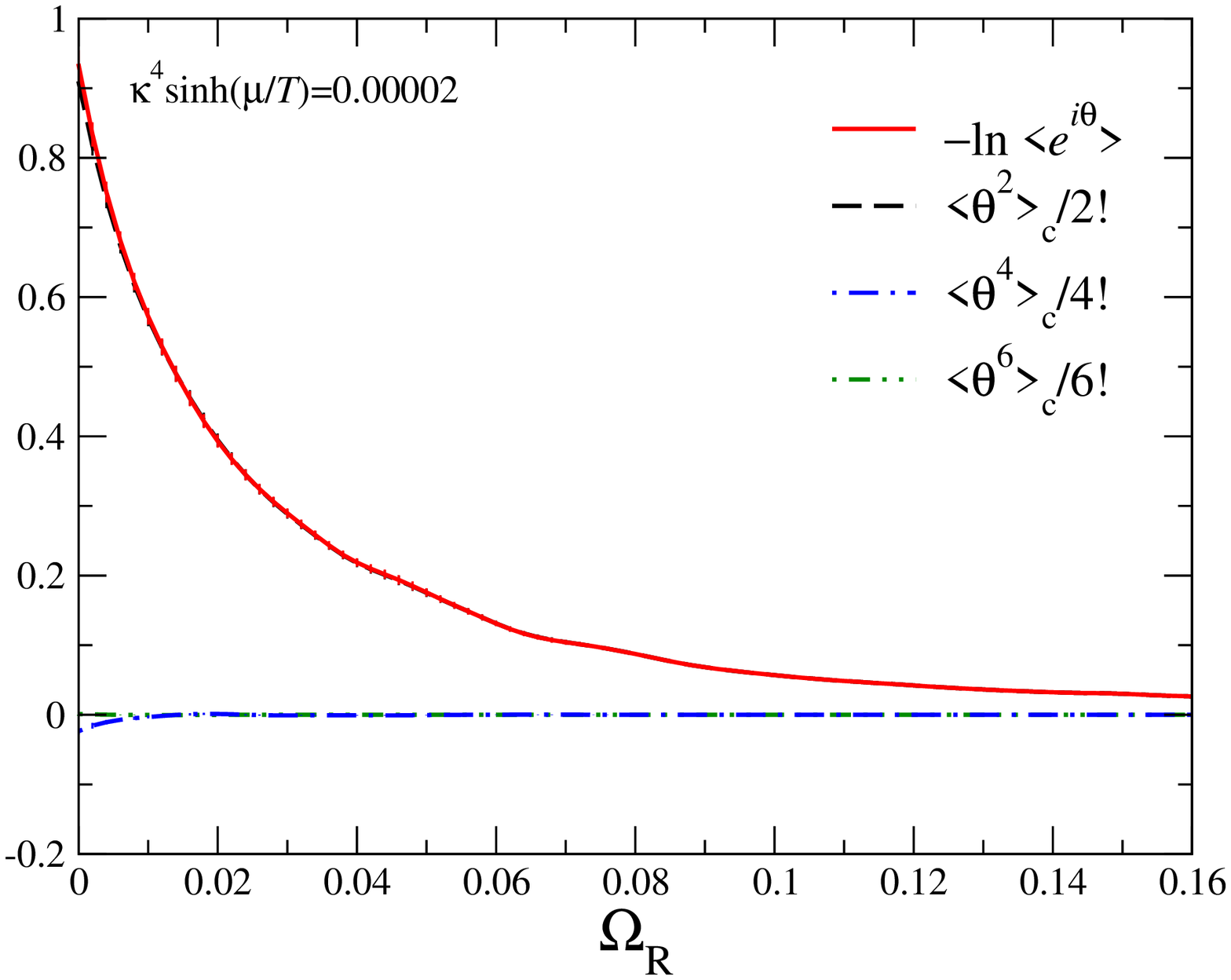}
\hspace{2mm}
\includegraphics[width=75mm]{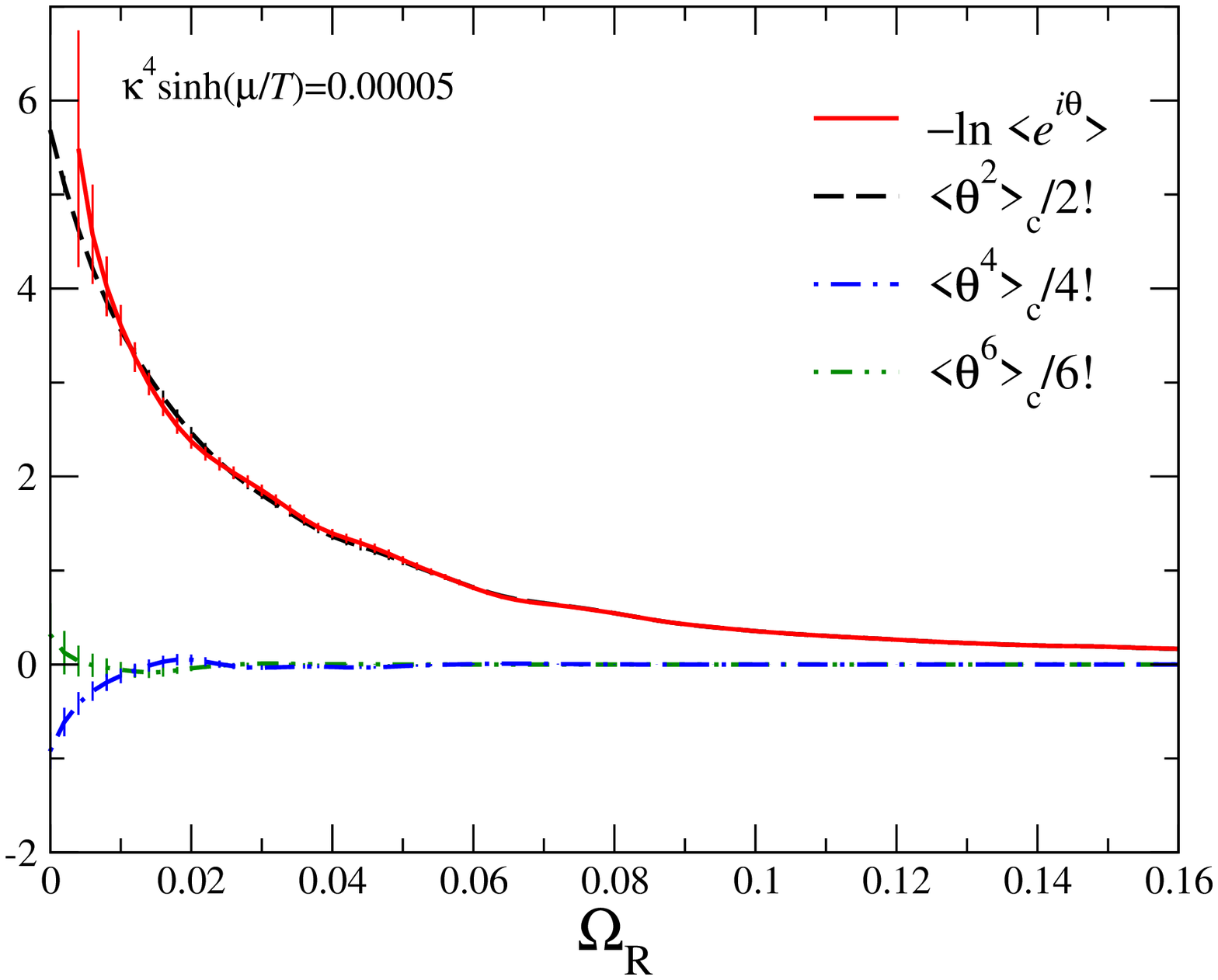}
}
\caption{Exponent of the average phase factor, $-\ln \langle e^{i\hat\theta} \rangle$, compared with the contributions from the second-, fourth-, and sixth-order cumulants. 
The expectation values are calculated at $\beta^*=5.69$ and $\kappa^4 \sinh (\mu/T) \approx 0.00002$ (left) or $0.00005$ (right) in $N_{\rm f}=2$ QCD with fixed $\Omega_{\rm R}$.
In both cases, $-\ln \langle e^{i\hat\theta} \rangle$ is almost indistinguishable with $\langle \hat\theta^2 \rangle_c / 2!$.}
\label{fig:phase}
\end{figure}

\begin{figure}[t] 
\centerline{
\includegraphics[width=75mm]{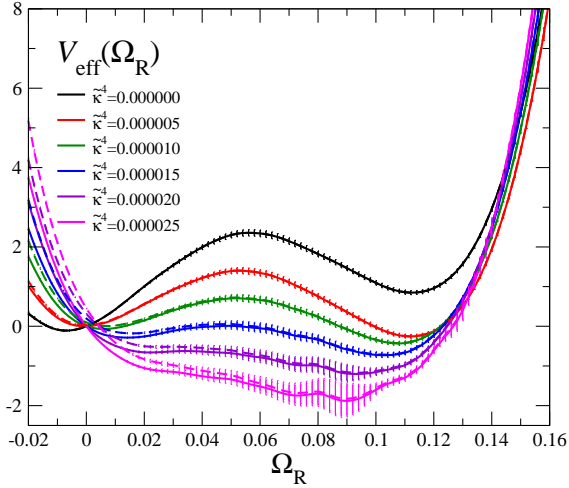}
}
\caption{
Effective potential of the Polyakov line in $N_{\rm f}=2$ heavy quark QCD on a $24^3 \times 4$ lattice.
Solid curves are $V_{\rm eff} (\Omega_{\rm R})$ in the phase-quenched approximation at various ${\tilde\kappa}^4 = \kappa^4 \cosh (\mu/T)$ where $\beta$ is adjusted to the transition point at each $\kappa$.
Dashed curves are the results for $V_{\rm eff} (\Omega_{\rm R})$ including the effects of the phase factor at $\mu/T=\infty$.
$V_{\rm eff} (\Omega_{\rm R})$ at finite $\mu/T$ are between the solid and dashed curves.
}
\label{fig:orhist}
\end{figure}

We now calculate the complex phase factor.
If $e^{i \hat{\theta}}$ changes its sign frequently, the statistical error becomes larger than the expectation value, causing the sign problem.
To avoid the sign problem, we evaluate the phase factor by the cumulant expansion method \cite{Ejiri:2007ga,whot10}: 
\begin{eqnarray}
\left\langle e^{i \hat\theta} \right\rangle_{\Omega_{\rm R}; (\beta^*, \vec0,\vec0)} 
= \exp \left[ \sum_{n=1}^{\infty} i^n \frac{\langle \hat\theta^n \rangle_c}{n!} \right], 
\label{eq:cum}
\end{eqnarray}
where $\langle \hat\theta^n \rangle_c$ is the $n$th-order cumulant with fixed $\Omega_{\rm R}$:
$
\langle \hat\theta^2 \rangle_c 
= \langle \hat\theta^2 \rangle_{\Omega_{\rm R}; (\beta^*, \vec0,\vec0)} 
$, 
$
\langle \hat\theta^4 \rangle_c 
= \langle \hat\theta^4 \rangle_{\Omega_{\rm R}; (\beta^*, \vec0,\vec0)} 
-3 \langle \hat\theta^2 \rangle_{\Omega_{\rm R}; (\beta^*, \vec0,\vec0)} ^2
$, 
$
\langle \hat\theta^6 \rangle_c 
= \langle \hat\theta^6 \rangle_{\Omega_{\rm R}; (\beta^*, \vec0,\vec0)} 
-15 \langle \hat\theta^4 \rangle_{\Omega_{\rm R}; (\beta^*, \vec0,\vec0)}  
\langle \hat\theta^2 \rangle_{\Omega_{\rm R}; (\beta^*, \vec0,\vec0)} 
+30 \langle \hat\theta^2 \rangle_{\Omega_{\rm R}; (\beta^*, \vec0,\vec0)}^3 
$, etc.
A key observation is that $\langle \hat\theta^n \rangle_c =0$ for any odd $n$
due to the symmetry under $\hat\theta \rightarrow -\hat\theta$. 
This implies that $\langle e^{i \hat\theta} \rangle$ is real and positive. 
Therefore, the sign problem is resolved if the cumulant expansion converges. 

The most convergent case, in which the leading term $\langle \hat\theta^2\rangle_c$ dominates in the expansion, corresponds to the case of Gaussian distribution. 
The distribution of the complex phase was found to be quite close to Gaussian in the light quark region of QCD both with Wilson-type and staggered-type improved quarks  up to moderate values of $\mu$, provided that the phase $\hat{\theta}$ is appropriately defined \cite{nakagawa,Ejiri:2007ga,whot10,splitt}.
See also Ref.~\cite{GMS13} for a recent study of the Gaussian dominance.
In the present case, $\hat\theta$ is given in terms of $\hat\Omega_{\rm I}$,
\begin{eqnarray} 
\left\langle \hat{\theta}^{2n} \right\rangle_c = \left( 3\times 2^{N_t+2} N_s^3 
\, q  
\right)^{2n} 
\left\langle \hat\Omega_{\rm I}^{2n} \right\rangle_c 
\label{eq:theta2nc}
\end{eqnarray} 
with $q = \sum_{f=1}^{N_{\rm f}} \kappa_f^{N_t} \sinh (\mu_f /T)$.
Figure~\ref{fig:OmegaI} shows our result for the distribution of $\Omega_{\rm I}$ for fixed $\Omega_{\rm R}$, obtained at $\vec\kappa=0$ and $\beta$ at the transition point.
The delta function is approximated by $\delta(x) \approx \exp[-(x/\Delta)^2]/(\Delta \sqrt{\pi}) $ with $\Delta=0.005$ for $\Omega_{\rm R}$ and $\Delta=0.001$ for $\Omega_{\rm I}$. 
The dashed lines in this plot are Gaussian functions fitted to the data at $\Omega_{\rm R}=0.0$ and $0.08$. 
We find that the distribution can be well approximated by a Gaussian function at $\Omega_{\rm R} \ge 0$. 
The deviation from Gaussian at $\Omega_{\rm R} < 0$ is due to the Z(3) symmetry at $\kappa=0$. 
As shown in Fig.~\ref{fig:compol}, at $\kappa>0$, the Z(3) symmetry is violated and the branch at $\Omega_{\rm R} > 0$ on the real axis is selected.
Therefore, in the determination of $\kappa_{\rm cp}$, only the $\Omega_{\rm R} > 0$ region is relevant.

We also note that $\hat\theta$ is given in terms of the spatially local operator $\hat\Omega_{\rm I}(\vec{x})$ in Eq.~(\ref{eq:thetaOmegaI}), where $\vec{x}$ is the spatial coordinate.
$\hat\Omega_{\rm I}(\vec{x})$ has finite correlation length given by the inverse electric screening mass \cite{eswhot10}.
We can thus decompose $\hat\theta$ into contributions from approximately uncorrelated spatial blocks, $\hat{\theta}= \sum_{x'} \hat{\theta}_{x'}$.
We then have \cite{whot10} 
\begin{eqnarray}
\left\langle e^{i\hat\theta} \right\rangle 
\approx \prod_{x'} \left\langle e^{i\hat\theta_{x'}} \right\rangle 
= \exp \left( \sum_{x'} \sum_n \frac{i^n}{n!} 
\left\langle \hat\theta_{x'}^n \right\rangle_c \right).
\end{eqnarray}
This implies the following.
First, $\langle \hat\theta^n \rangle_c$ is linearly proportional to the system volume for any $n$, in contrast to a naive expectation of (volume)$^n$ since $\hat\theta$ is proportional to the volume.
Thus, the range in which the cumulant expansion is applicable is independent of the volume, 
in spite of the fact that the sign problem becomes exponentially  serious with the volume. 
Second, when the system size is much larger than the correlation length, the distribution of $\theta/$volume tends to a Gaussian distribution according to the central limit theorem.
The property $\langle \hat\theta^n \rangle_c \propto$ the system volume at any $n$ can be understood by noting that it is a sufficient condition to have a well-defined effective potential 
$V_{\rm eff}(\mu) =V_{\rm eff}(\mu=0) - \ln \langle e^{i \hat{\theta}} \rangle = V_{\rm eff}(\mu=0) - \sum_n i^{n} \langle \hat{\theta}^{n} \rangle_c/n!$ around $\mu=0$ in the large volume limit, 
because $V_{\rm eff}$ is proportional to the volume. 

In Fig.~\ref{fig:phase}, we plot $\langle \hat\theta^n \rangle_c /n!$ in $N_{\rm f}=2$ QCD at $\beta^* =5.69$ and $\kappa^{N_t} \sinh (\mu/T)=0.00002$ (left) and 0.00005 (right).
The dashed, dot-dashed and two-dot-dashed lines are the results for $n=2$, 4, and 6, respectively.
The red solid line represents 
$- \ln \langle e^{i \hat{\theta}} \rangle_{\Omega_{\rm R}; (\beta^*, \vec0,\vec0)} \ [ = - \ln \langle \cos \hat{\theta} \rangle_{\Omega_{\rm R}; (\beta^*, \vec0,\vec0)} ]$, 
which is almost indistinguishable from the second order cumulant.
We find that the complex phase factor can be well described by the Gaussian approximation at these points.
The higher order contributions become visible at small $\Omega_{\rm R}$ at large $\mu$ as shown in the right panel of Fig.~\ref{fig:phase}. 
However, for the determination of the critical point, $\kappa_{\rm cp}^{N_t} \cosh (\mu/T) \approx 0.00002$ in the phase-quenched theory, the region at $\kappa^{N_t} \sinh (\mu/T) < 0.00002$ is important because $\sinh (\mu/T) < \cosh (\mu/T)$.
Thus, we can safely adopt the Gaussian approximation around the critical point in the heavy quark region at all values of $\mu/T$ including $\mu/T=\infty$.
See Appendix B for a discussion on the application range of the Gaussian approximation,
in which we estimate the parameter region of $\kappa^{N_t} \sinh (\mu/T)$ where the second order term dominates over the higher order terms.

Our results for the effective potential in $N_{\rm f}=2$ QCD including the effect of the complex phase factor are shown in Fig.~\ref{fig:orhist} for various ${\tilde\kappa}^{N_t} = \kappa^{N_t} \cosh (\mu/T)$ at $N_t=4$.
The solid curves are $V_{\rm eff} (\Omega_{\rm R})$ in the phase-quenched approximation at various ${\tilde\kappa}^{N_t}$, 
which are identical to  the $V_{\rm eff} (\Omega_{\rm R})$ at $\mu=0$ for each ${\tilde\kappa}^{N_t} = \kappa^{N_t}$ shown in the left panel of Fig.~\ref{fig:oabhist}.
The dashed curves are $V_{\rm eff} (\Omega_{\rm R})$ taking account of the contribution of the phase factor according to Eq.(\ref{eq:plhvdist}), using the Gaussian approximation for $ \langle e^{i \hat{\theta}} \rangle_{\Omega_{\rm R}; (\beta^*, \vec0,\vec0)} $.
To estimate the upper bound of the phase factor effects, we have set $\kappa^{N_t} \sinh(\mu/T) = \kappa^{N_t} \cosh(\mu/T)$ in $\hat\theta$, corresponding to the case of $\mu/T=\infty$.
Because $\sinh (\mu/T) < \cosh (\mu/T)$ at $\mu/T < \infty$, 
$V_{\rm eff} (\Omega_{\rm R})$ at finite $\mu/T$ is between the solid and dashed curves. 

We find that the contribution from the phase factor is quite small except at small $\Omega_{\rm R}$.
For the determination of the critical point, the shape of $V_{\rm eff}$ around the point where the two minima merges is important. 
Around the values of $\Omega_{\rm R}$ and $\tilde\kappa$ relevant to the determination of the critical point, the dashed curves are quite close to the solid curves. 
This means that the contribution from the complex phase to the location of the critical point is quite small on our $24^3 \times 4$ lattice. 
The critical point $\kappa_{\rm cp}(\mu)$ in the full theory can be well approximated to that in the phase-quenched theory up to $\mu/T=\infty$,
\begin{equation}
\kappa_{\rm cp}(\mu) \approx \kappa_{\rm cp}^I (\mu) ,
\label{eq:full}
\end{equation}
with $\kappa_{\rm cp}^I (\mu)$ given by Eq.~(\ref{eq:isospin}).

\subsection{Critical surface in 2+1 flavor QCD}
\label{Sec:2+1flavor}

\begin{figure}[tb]
\centerline{
\includegraphics[width=76mm]{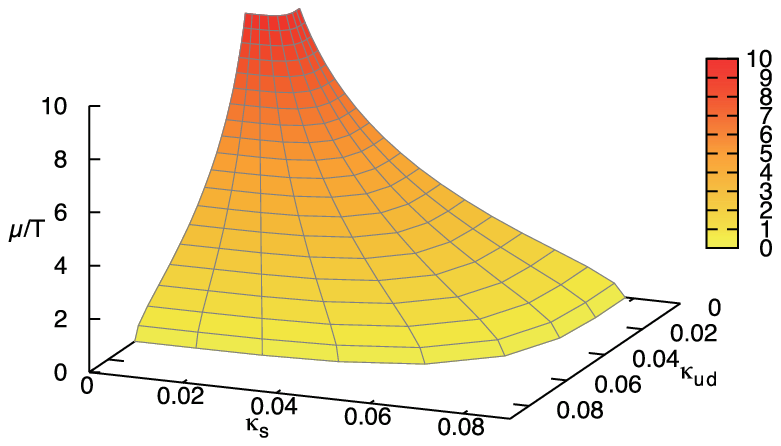}
\includegraphics[width=76mm]{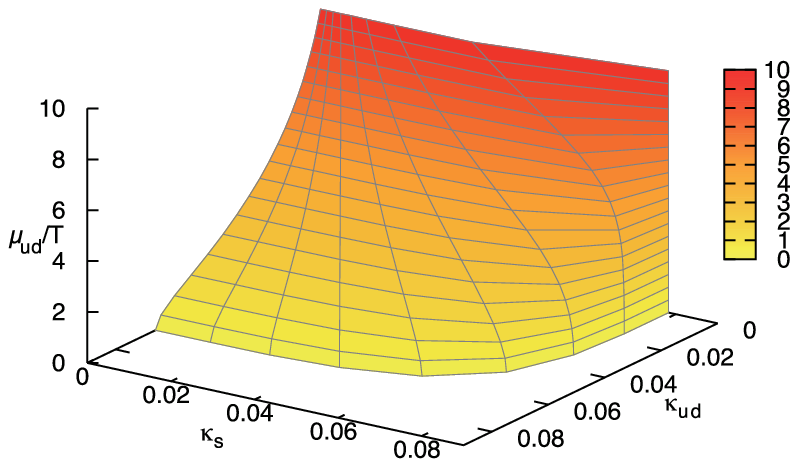}
}
\caption{Critical surface separating the first order transition and crossover regions in the heavy quark region.
Left: 
The case $\mu_{u} = \mu_{d} = \mu_{s} \equiv \mu$.
Right: 
The case that may be realized in heavy ion collisions: $\mu_{u} = \mu_{d} \equiv \mu_{ud}$ and $\mu_{s} = 0$.
}
\label{fig:crtsur}
\end{figure}

It is straightforward to extend the discussions to the case of nondegenerate heavy quark QCD using Eq.(\ref{eq:plhvdist}).
As discussed in the previous subsection, the effects of the complex phase factor are negligible in the determination of the critical point. 
Therefore, the critical point in nondegenerate heavy quark QCD is given well by 
\begin{equation}
h(\vec\kappa_{\rm cp},\vec\mu_{\rm cp}) = 2 [\kappa_{\rm cp}^{N_{\rm f}=2}(0) ]^{N_t} ,
\end{equation}
where $h$ is defined by Eq.~(\ref{eq:h}) and $\kappa_{\rm cp}^{N_{\rm f}=2}(0)=0.0658(3)(^{+4}_{-11})$ at $N_t=4$ 
\cite{PRD84_054502}.
In particular, the critical point in heavy-quark $N_{\rm f}=2+1$ QCD is given by 
\begin{equation}
2 \, \kappa_{\rm ud}^{N_t} (\mu_{\rm ud}, \mu_{\rm s}) 
\cosh \left( \frac{\mu_{\rm ud}}{T} \right) 
+ \kappa_{\rm s}^{N_t} (\mu_{\rm ud}, \mu_{\rm s}) 
\cosh \left( \frac{\mu_{\rm s}}{T} \right) 
= 2 [\kappa_{\rm cp}^{N_{\rm f}=2}(0) ]^{N_t}  
\end{equation}
in the coupling parameter  space $(\kappa_{\rm ud}, \kappa_{\rm s},\mu_{\rm ud},\mu_{\rm s})$.
The critical surfaces for the cases (a) $\mu_{\rm ud}/T=\mu_{\rm s}/T\ne0$ (left) and (b) $\mu_{\rm ud}/T\ne0$ with $\mu_{\rm s}/T=0$ are shown in the left and right panels of Fig.~\ref{fig:crtsur}, respectively.
In realistic heavy-ion collision experiments, because the strange quarks are hardly prepared in the environment, $\mu_{\rm s}\approx0$ is realized.
This result is consistent with that of the effective theory in Ref.~\cite{Fromm12}. 
We note that these results are obtained at $N_t=4$. 
For a precise prediction to be compared with the real world, 
it is important to study the lattice cutoff dependence ($N_t$-dependence) of the critical surface. 
We leave this as a future work.

\section{Effective potential for $(P,\Omega_{\rm R})$}
\label{sec:popot}

In the previous sections, we have studied effective potentials for $P$ and $\Omega_{\rm R}$ separately.
Because the gauge action and quark determinant Eq.~(\ref{eq:detM}) are given by the plaquette and Polyakov line in the heavy quark region, we may instead consider an effective potential for $(P,\Omega_{\rm R})$ simultaneously.
In heavy-quark QCD, $P$ and $\Omega_{\rm R}$ represent the freedom of gauge and quark free energies, respectively.
Therefore, we expect that the histogram $w(P,\Omega_{\rm R})$ and its effective potential $V_{\rm eff}(P,\Omega_{\rm R})$ are sensitive to the phase transition. 

To the leading order of the hopping parameter expansion, the histogram of $(P,\Omega_{\rm R})$ is given by
\begin{eqnarray}
w(P, \Omega_{\rm R}; \beta, \vec\kappa, \vec\mu ) 
&=& \int {\cal D} U \, \delta (P - \hat{P}) \,
\delta (\Omega_{\rm R} - \hat{\Omega}_{\rm R}) \ 
e^{-S_g (\beta ) }\ \prod_{f=1}^{N_{\rm f}} \det M (\kappa_f, \mu_f) 
\nonumber \\
&& \hspace{-24mm} \approx \int \! \! {\cal D} U \, \delta ( P - \hat{P} ) \,
\delta ( \Omega_{\rm R} - \hat{\Omega}_{\rm R} ) \, e^{6N_{\rm site} \beta \hat{P} } 
\nonumber \\
&& \hspace{-14mm} \times 
e^{ \left[ 288N_{\rm site} 
\sum_{f=1}^{N_{\rm f}}  \kappa_f^4 \hat{P} 
+ 3 \times 2^{N_t+2} N_s^3 \sum_{f=1}^{N_{\rm f}} \kappa_f^{N_t} 
\left\{ \cosh \left( \frac{\mu_f}{T} \right) \hat{\Omega}_{\rm R} 
+i \sinh \left( \frac{\mu_f}{T} \right) \hat{\Omega}_{\rm I} \right\} \right] } 
\nonumber \\
&& \hspace{-24mm} = e^{ 6N_{\rm site} (\beta^* -\beta_0) \, P 
+ 3 \times 2^{N_t+2} N_s^3 h \, \Omega_{\rm R} } \,
w(P, \Omega_{\rm R}; \beta_0, \vec0, \vec0) \times
\left\langle e^{i \hat{\theta} } \right\rangle_{P, \Omega_{\rm R};(\beta_0,\vec0,\vec0)} 
\label{eq:ppdist}
\end{eqnarray}
where $\beta^*$ and $h$ are given by Eqs.~(\ref{eq:betastar}) and (\ref{eq:h}).
The histogram is thus factored into the phase-quenched part and the complex phase factor part $\langle e^{i \hat{\theta}} \rangle$.
The $\beta$ and $\vec\kappa$ dependences in the phase-quenched part are quite simple.
Although the statistical quality of $w(P,\Omega_{\rm R})$ is worse than that for $w(P)$ or $w(\Omega_{\rm R})$, 
this simple dependence on the coupling parameters may help us to investigate the phase structure in a wide range of the coupling parameter space.

\begin{figure}[tb]
\vspace{-9mm}
\centerline{
\includegraphics[width=59mm]{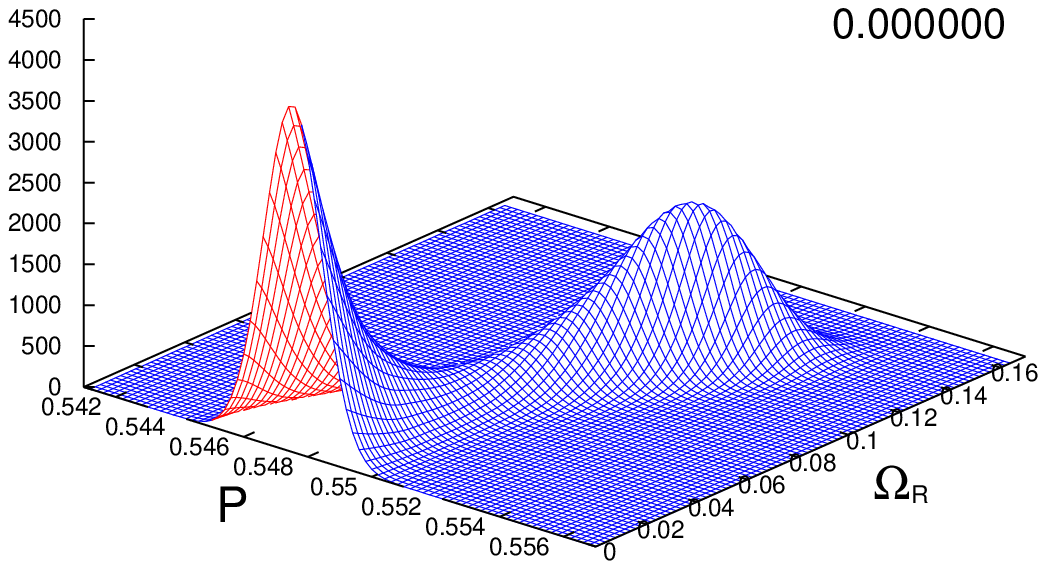}
\hspace{-11mm}
\includegraphics[width=59mm]{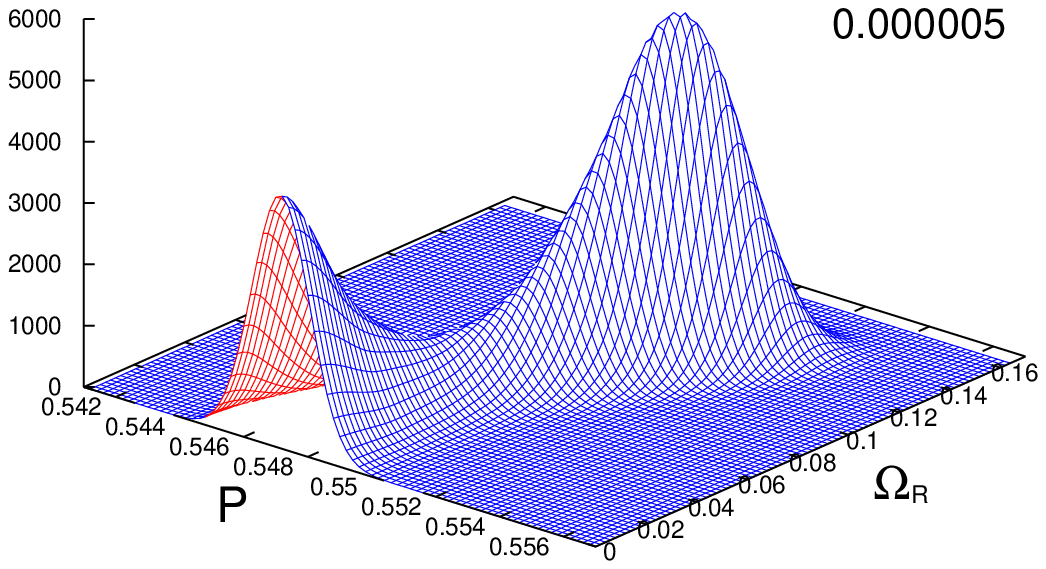}
\hspace{-11mm}
\includegraphics[width=59mm]{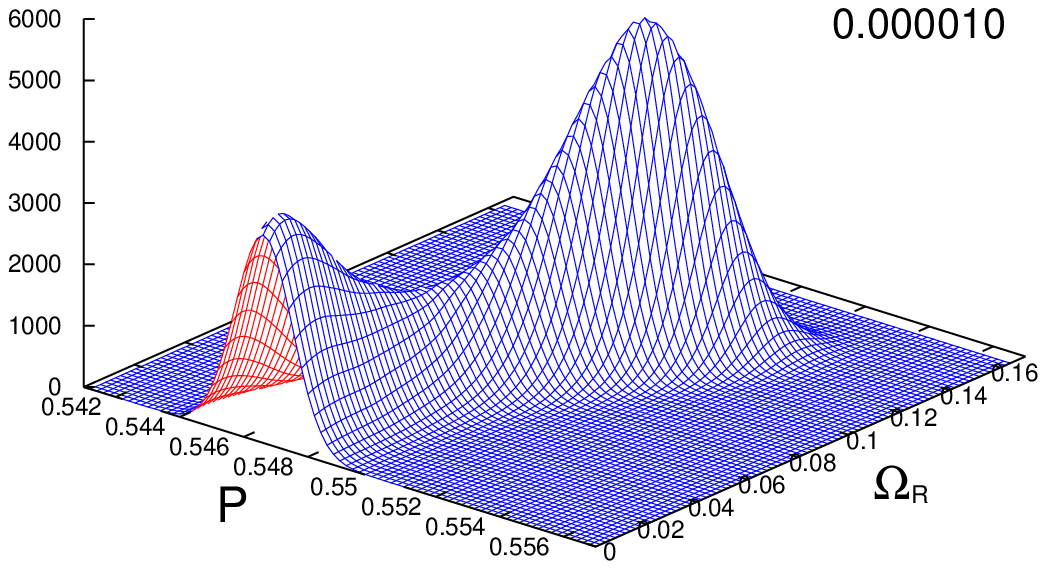}
}
\vspace{-12mm}
\centerline{
\includegraphics[width=59mm]{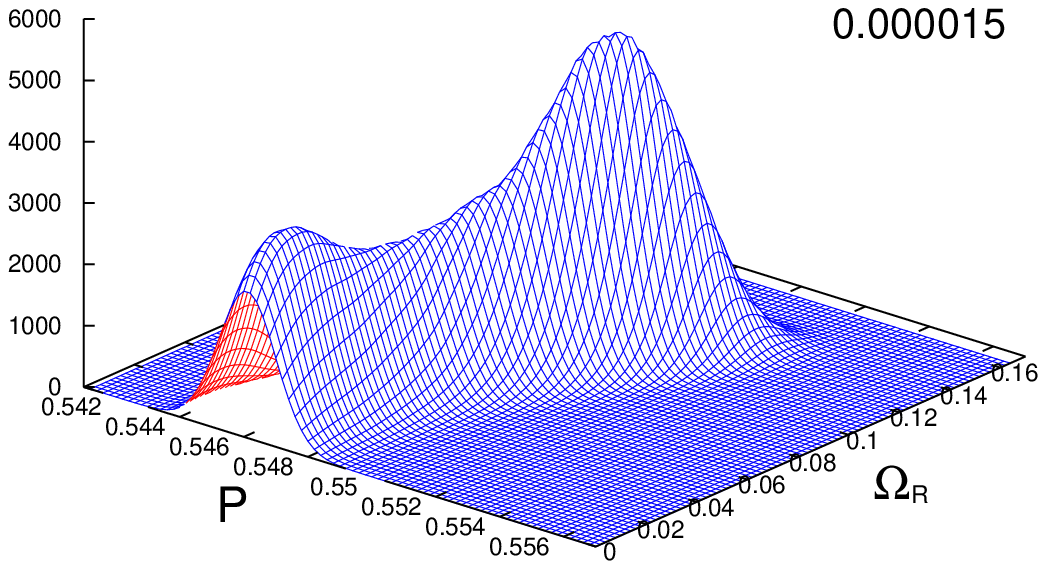}
\hspace{-11mm}
\includegraphics[width=59mm]{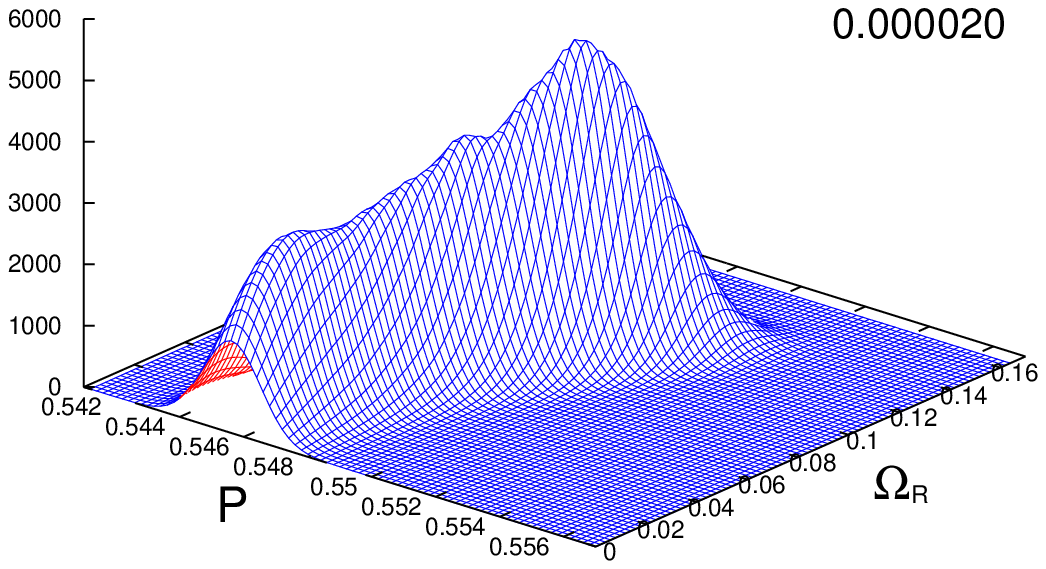}
\hspace{-11mm}
\includegraphics[width=59mm]{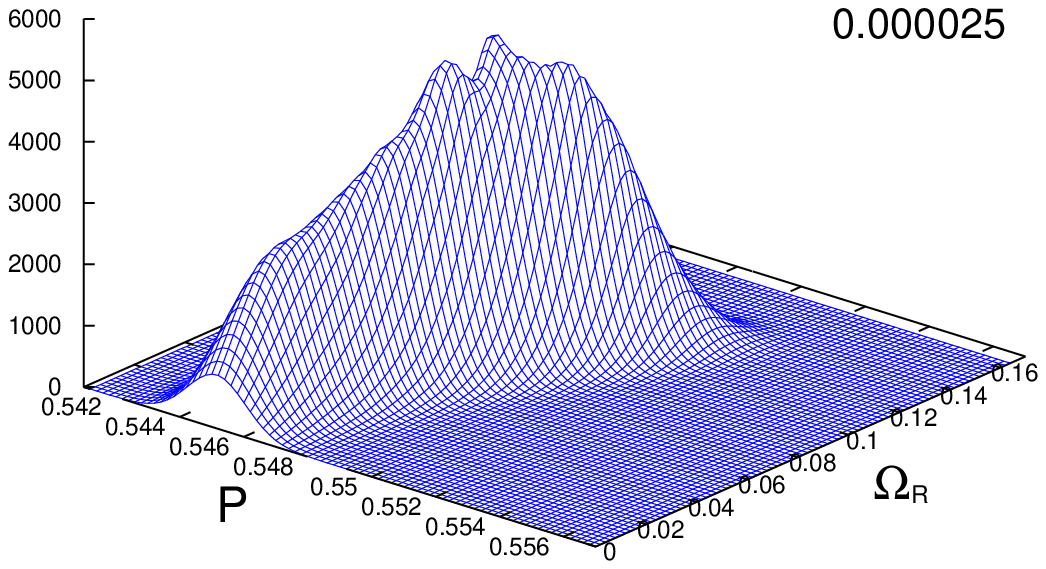}
}
\vspace{-2mm}
\caption{
Histogram  $w(P, \Omega_{\rm R};\beta,\kappa,0)$ as function of $\kappa^4$ in $N_{\rm f}=2$ QCD.
The value of $\kappa^4$ is shown in the upper right corner of each plot.
$\beta$ is adjusted to the peak position of the Polyakov line susceptibility at each $\kappa$. 
}
\label{fig:plqpl}
\end{figure}

\subsection{Effective potential and critical point at zero density}
\label{sec:derpot0}

Let us first study the case $\vec\mu=0$ in which the complex phase factor is absent in Eq.~(\ref{eq:ppdist}). 
The histogram $w(P, \Omega_{\rm R};\beta,\vec\kappa,\vec0)$ is shown in Fig.~\ref{fig:plqpl} in the case of degenerate $N_{\rm f}=2$ QCD, 
where $\beta$ is adjusted to the peak position of the Polyakov line susceptibility at each $\kappa$. 
Data are taken from the pure-gauge configurations generated on a $24^3\times 4$ lattice \cite{PRD84_054502}. 
Using the multi-point reweighting method, we combine data at five $\beta$ points in the range $\beta = 5.68$--5.70. 
The total number of configurations is 1,800,000. 
With these configurations, the plaquette and the Polyakov line are distributed in the range $0.543 \simle P \simle 0.556$ and $| \Omega | \simle 0.15$.
To evaluate the pure-gauge histogram $w(P, \Omega_{\rm R};\beta,\vec0,\vec0)$, we approximate the delta function by a Gaussian function $\delta(x) \approx \exp [ -(x/\Delta)^2 ] / (\Delta \sqrt{\pi})$, where $\Delta =0.0005$ for $P$ and 0.005 for $\Omega_{\rm R}$ by consulting the statistical stability of the final results, and average over the Z(3) rotations of $\Omega$. 
Then, $w(P, \Omega_{\rm R};\beta,\vec\kappa,\vec0)$ at finite $\kappa$ are computed by the reweighing formula.
We see that the two peaks at low $\kappa$ merge into a single peak as we increase $\kappa$, suggesting a critical point beyond which the first order transition turns into a crossover.

\begin{figure}[t]
   \begin{minipage}{80mm}
   \includegraphics[width=80mm]{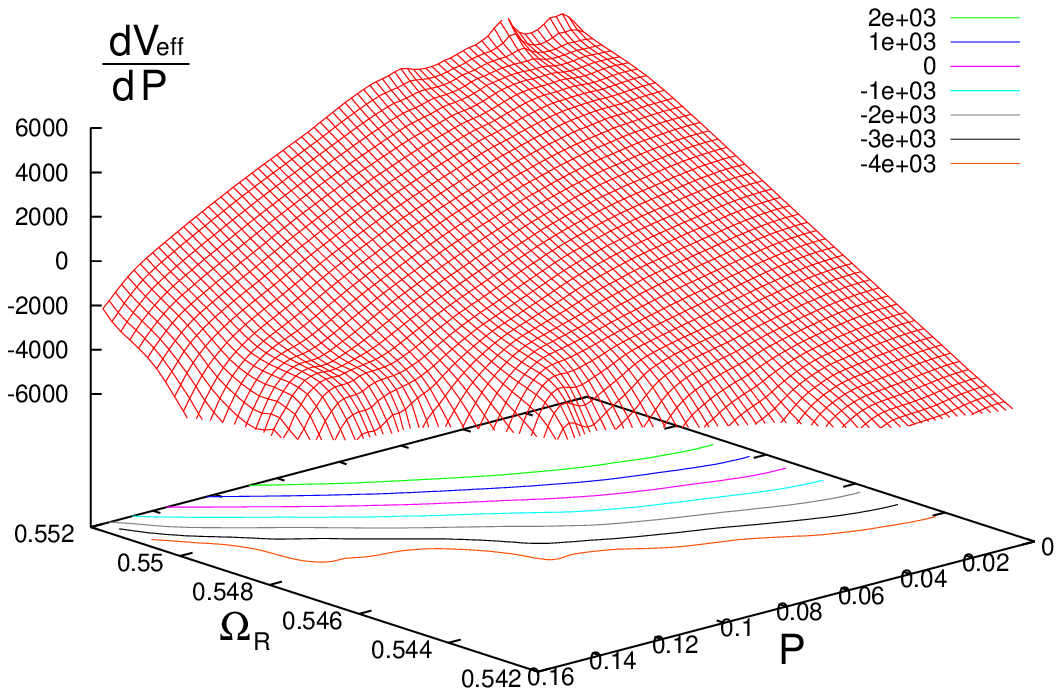} 
   \end{minipage}
   \hspace{-4mm}
   \begin{minipage}{80mm}
   \includegraphics[width=80mm]{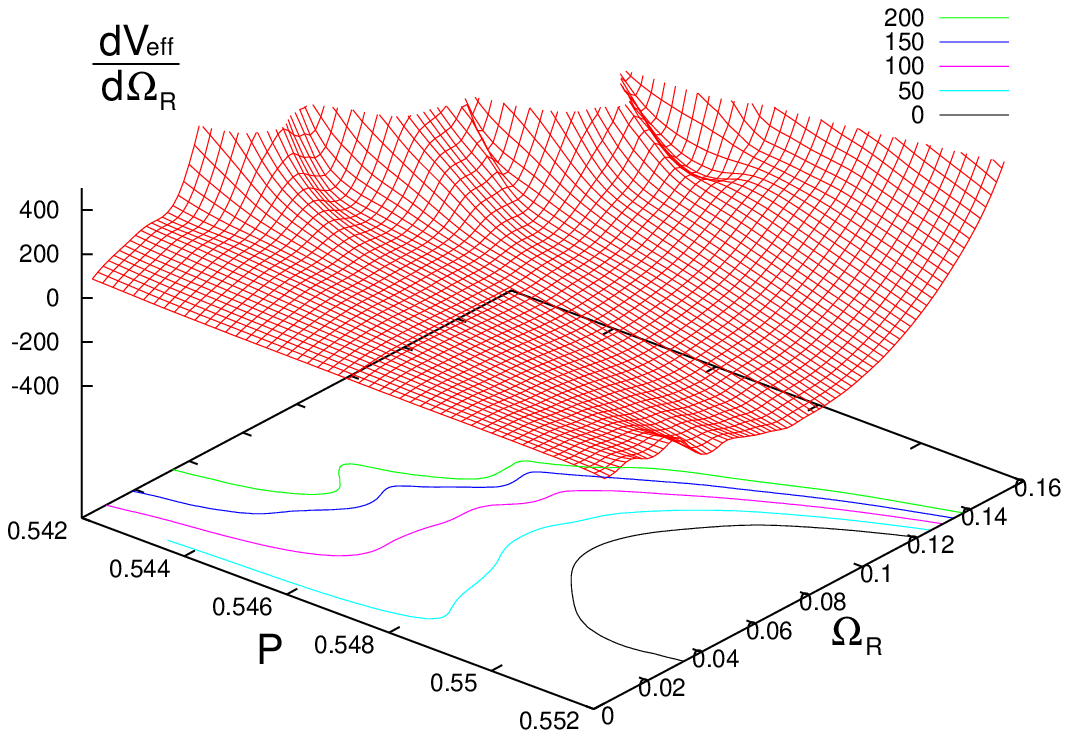} 
   \end{minipage}
   \caption{
   $\partial V_0/\partial P$ (left) and $\partial V_0/\partial \Omega_{\rm R}$ (right) at $\beta_0=5.69$ as functions of $(P,\Omega_{\rm R})$.
   Their contour curves are given at the bottom of the plots.
   }
   \label{fig:dV0dP_dV0dOmega}
\end{figure}

We now define the effective potential $V_{\rm eff}(P, \Omega_{\rm R}; \beta, \vec\kappa,\vec{0}) = - \ln w(P, \Omega_{\rm R}; \beta, \vec\kappa,\vec{0})$.
The peak position of the histogram corresponds to the point where the curves $\partial V_{\rm eff}/\partial P = 0$ and $\partial V_{\rm eff}/\partial \Omega_{\rm R} = 0$ intersect in the $(P, \Omega_{\rm R})$ plane.
When we have only one peak of the histogram, the two curves intersect at just one point.
On the other hand, when we have two peaks of the histogram, the two curves intersect at three points, corresponding to two peaks and one saddle point between the peaks.
Therefore, from a merger of these intersection points into one, we can detect the critical point where the first order transition turns into a crossover.

According to Eq.~(\ref{eq:ppdist}) at $\vec\mu=0$, the derivatives of $V_{\rm eff}(P, \Omega_{\rm R}; \beta, \vec\kappa,\vec{0})$ have simple dependences on $\beta$ and $\vec\kappa$,
\begin{eqnarray}
  \frac{\partial V_{\rm eff}}{\partial P}(P, \Omega_{\rm R}; \beta, \vec\kappa,\vec0)
  &=& \frac{\partial V_0}{\partial P}(P, \Omega_{\rm R}; \beta_0)  - 6N_{\rm site} \left(\beta^* -\beta_0 \right) ,
  \label{eq:dV0dP}
\\
  \frac{\partial V_{\rm eff}}{\partial \Omega_{\rm R}}(P, \Omega_{\rm R};\backslash\hspace{-2mm}\beta,\vec\kappa,\vec0) 
  &=& \frac{\partial V_0}{\partial \Omega_{\rm R}}(P, \Omega_{\rm R};\backslash\hspace{-2mm}\beta_0)  - 3\times2^{N_t+2} N_s^3 \, h ,
  \label{eq:dV0dOmega}
\end{eqnarray}
where $V_0(P, \Omega_{\rm R};\beta_0) = V_{\rm eff}(P, \Omega_{\rm R};\beta_0,\vec0,\vec0)$  is the effective potential in the heavy quark limit and $h = \sum_{f=1}^{N_{\rm f}} \kappa_f^{N_t} $ at $\vec\mu=0$.
Note that $\partial V_{\rm eff}/\partial \Omega_{\rm R}$ and $\partial V_{0}/\partial \Omega_{\rm R}$ are independent of $\beta$ and $\beta_0$,
while $\partial V_{\rm eff}/\partial P$ depends on $\beta$ and $\vec\kappa$ only through $\beta^*$. 
From these equations, we find that, in the derivatives of the effective potential, 
(i) $\beta$ and $\vec\kappa$ dependences exist only in the additive constant shifts, and 
(ii) besides these constant shifts, $P$ and $\Omega_{\rm R}$ dependences appear only in the derivatives of $V_0$ which do not depend on $\beta$ and $\vec\kappa$. 
Therefore, a curve for $\partial V_{\rm eff}/\partial P = 0$ at a $\beta^*$ point is independent of $\vec\kappa$ and is given by a contour of $\partial V_0 / \partial P$,
while a curve for $\partial V_{\rm eff}/\partial \Omega_{\rm R} = 0$ at some $\vec\kappa$ does not depend on $\beta$ and is given by a contour of $\partial V_0 / \partial \Omega_{\rm R}$.

\begin{figure}[t]
\vspace{-4mm}
\centerline{
\includegraphics[width=90mm]{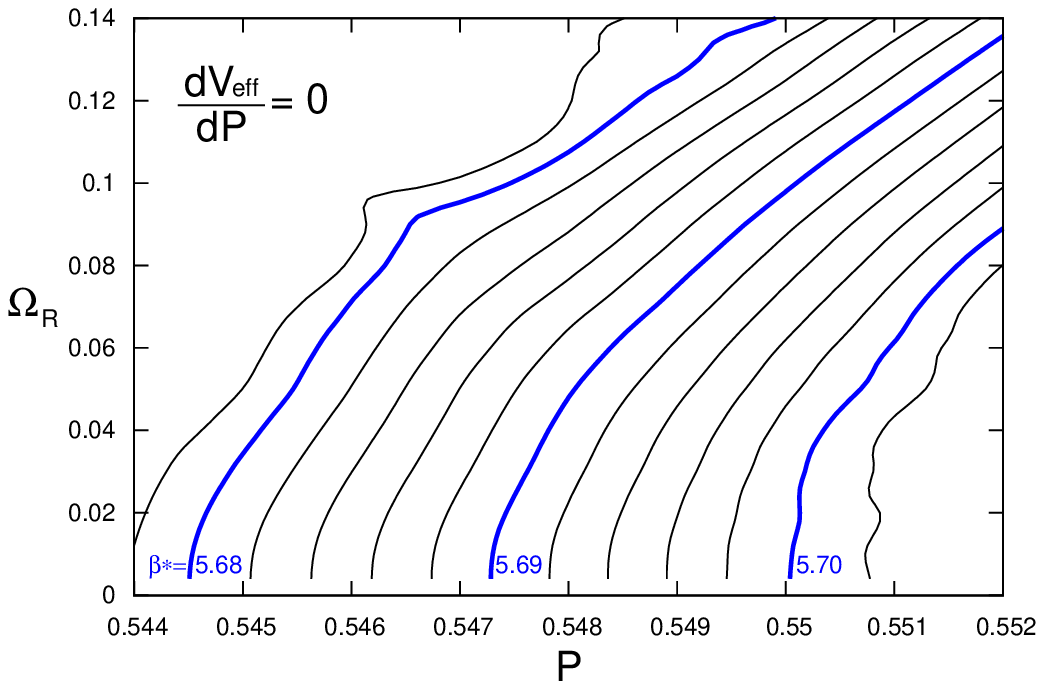}
\hspace{-12mm}
\includegraphics[width=90mm]{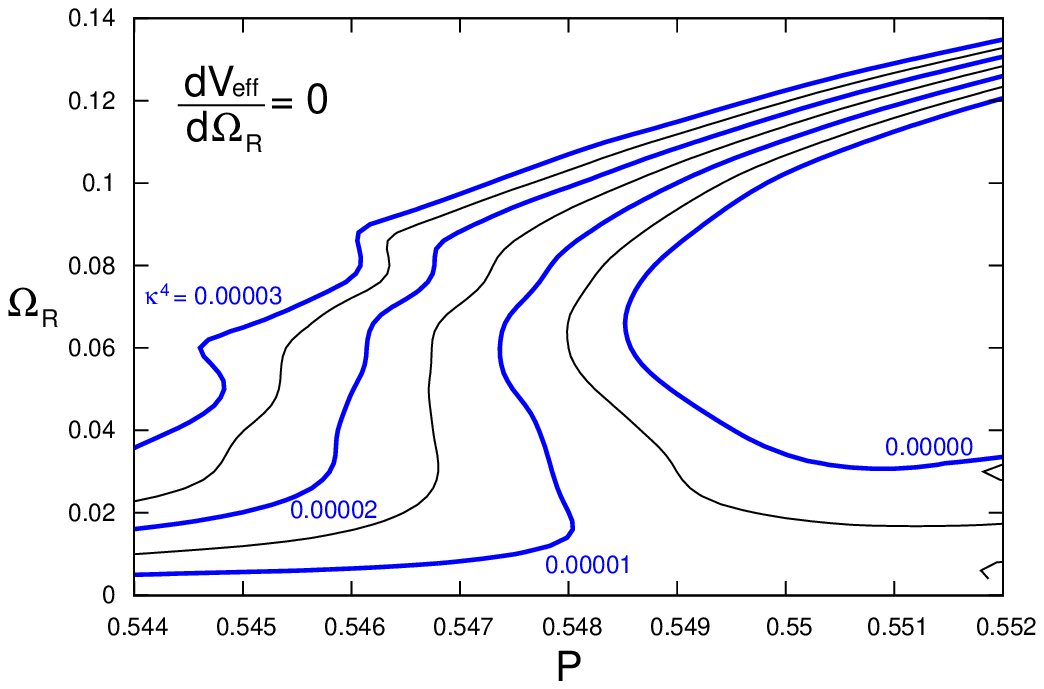}
}
\vspace{-4mm}
\caption{
Curves of $\partial V_{\rm eff}/\partial P = 0$ (left) and  $\partial V_{\rm eff}/\partial \Omega_{\rm R} = 0$ (right) at $\vec\mu=0$, 
Values of $\beta^*$ (left) and $\kappa^4$ (right) 
are for the case of $N_{\rm f}=2$ QCD. 
The curves are identical to the contour curves of $\partial V_0/\partial P$ and $\partial V_0/\partial \Omega_{\rm R}$ shown in Fig.~\ref{fig:dV0dP_dV0dOmega}. 
}
   \label{fig:contdvdpdO}
\end{figure}

\begin{figure}[t]
\vspace{-4mm}
\centerline{
\includegraphics[width=90mm]{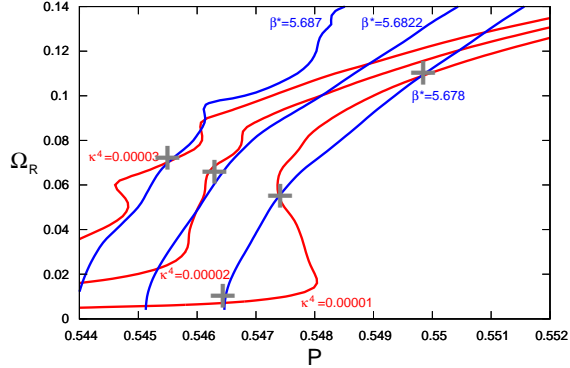}
}
\vspace{-4mm}
\caption{$\partial V_{\rm eff}/\partial P =0$ (blue curves) and $\partial V_{\rm eff}/\partial \Omega_{\rm R} =0$ (red curves) shown in Fig.~\ref{fig:contdvdpdO}, around the critical point. 
}
   \label{fig:cont_mu0}
\end{figure}

In Fig.~\ref{fig:dV0dP_dV0dOmega}, we plot $\partial V_0/\partial P$ and $\partial V_0/\partial \Omega_{\rm R}$ at $\beta_0=5.69$ as functions of $(P,\Omega_{\rm R})$. 
We calculate the derivatives by fitting the data between $x-\epsilon/2$ and $x+\epsilon/2$ 
by a linear function in $P$ and $\Omega_{\rm R}$, where $\epsilon = 0.0016$ for $P$ and 0.016
for $\Omega_{\rm R}$.
The contour curves of $\partial V_0/\partial P$ and $\partial V_0/\partial \Omega_{\rm R}$ 
can be viewed as the curves $\partial V_{\rm eff}/\partial P = 0$ and  $\partial V_{\rm eff}/\partial \Omega_{\rm R} = 0$ at different $(\beta,\vec\kappa)$,
as shown in Fig.~\ref{fig:contdvdpdO}.
The values of $\beta^*$ and $\kappa^4$ in this figure are for the case of $N_{\rm f}=2$ QCD. 
Recall that we combined configurations generated at $\beta =5.68$ -- $5.70$ with $\kappa=0$ using the multi-point reweighting method.
Therefore, the data in the region sandwiched between the curves $\beta^*=5.68$ and $5.70$ in the left panel of Fig.~\ref{fig:contdvdpdO} are trustworthy, while the regions around the upper left and lower right corners of this plot suffer from large fluctuations due to a poor statistics.%
\footnote{
In a preliminary version of this study \cite{Saito2}, we did not use the multi-point reweighting method and adopted a simple difference for the numerical derivative. The multi-point reweighting method enabled us to evaluate the derivatives in a wider range of $P$ and $\Omega_{\rm R}$ and to vary $\beta$ and $\kappa$ more systematically, while the statistical quality just around the critical point was not improved much.
}

We then overlay these curves in Fig.~\ref{fig:cont_mu0}.
At $\kappa^4=0.00001$ and $\beta^*=5.678$, we find three intersection points due to the letter S-like shape of the $\partial V_{\rm eff}/\partial \Omega_{\rm R}=0$ curve. 
The three intersection points correspond to two local minima and a saddle point of $V_{\rm eff}$. 
This means that we have two meta-stable states and thus have a first-order transition around this point. 
With increasing $\kappa$, the S shape becomes weaker, and eventually the three intersection points merge to one intersection point at $(P,\Omega_{\rm R}) \approx (0.546,0.06)$. 
This happens at the critical point around $\kappa^4 \approx 0.00002$ $(\kappa_{\rm cp} \approx 0.0669)$ and $\beta^* \approx 5.6822$ in $N_{\rm f}=2$ QCD.
Taking into account the ambiguities in identifying the shape of contours, we consider that these values for the critical point are roughly consistent with those obtained in Ref.~\cite{PRD84_054502} and Sec.~\ref{sec:pldist0} using $V_{\rm eff}(P)$ and $V_{\rm eff}(\Omega_{\rm R})$.

\subsection{Effective potential and critical point at finite density}
\label{sec:distmu}

We now turn on the chemical potential. 
We first study the phase-quenched theory.
The histogram $w(P,\Omega_{\rm R})$ in the phase-quenched theory is given by Eq.~(\ref{eq:ppdist}) with the complex phase factor suppressed.
Then, after absorbing the plaquette term into the gauge action by replacing $\beta$ by $\beta^*$, 
the only effect of $\vec\mu\ne0$ in the phase-quenched theory is to replace the hopping parameter
$\kappa_f$ in the theory at $\vec\mu=0$ by $\kappa_f \cosh^{1/N_t} (\mu_f/T)$. 
Therefore, the discussions in Sec.~\ref{sec:isospin} for $w(\Omega_{\rm R})$ are applicable to the present case too, 
and the critical point in the phase-quenched theory is given by Eq.~(\ref{eq:isospin}).

\begin{figure}[t] 
\centerline{
\includegraphics[width=8.5cm]{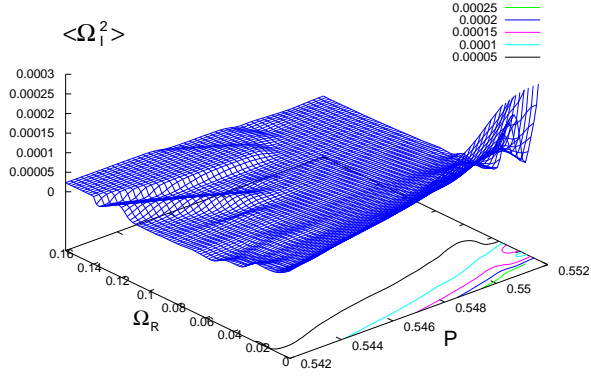}
}
\vspace{-5mm}
\caption{$\langle \hat\Omega_{\rm I}^2 \rangle_c$ with fixed $P$ and $\Omega_{\rm R}$.}
\label{fig:OmegaI2}
\end{figure}

\begin{figure}[t] 
\centerline{
\includegraphics[width=8.5cm]{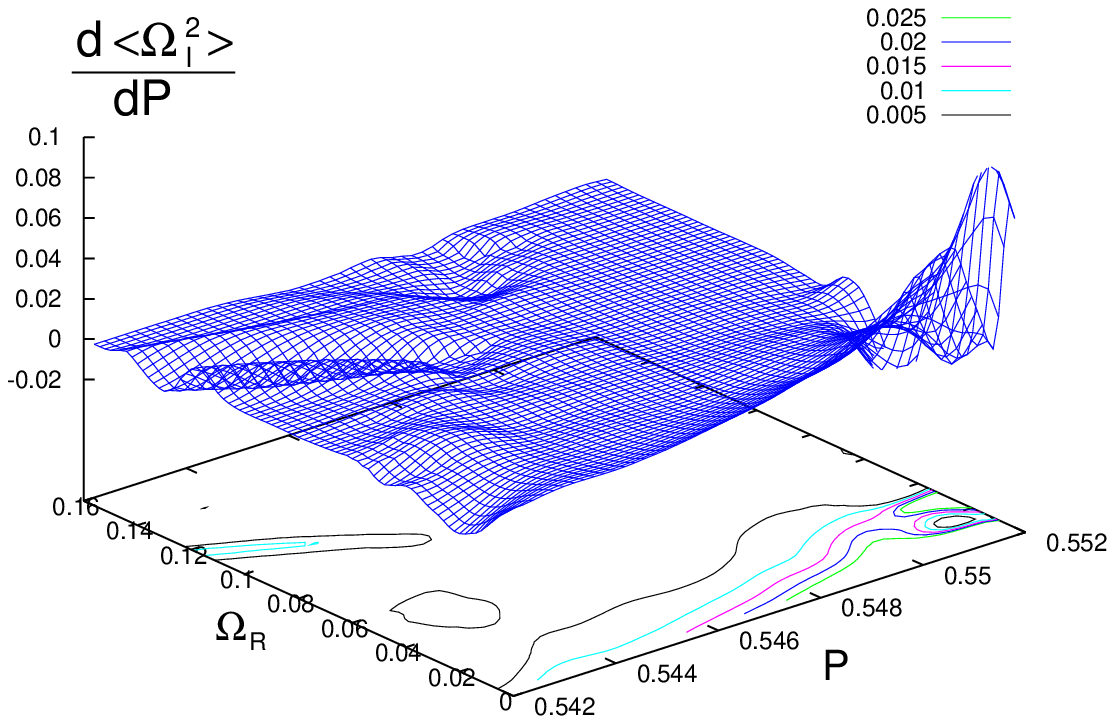} 
\hspace{-0.7cm}
\includegraphics[width=8.5cm]{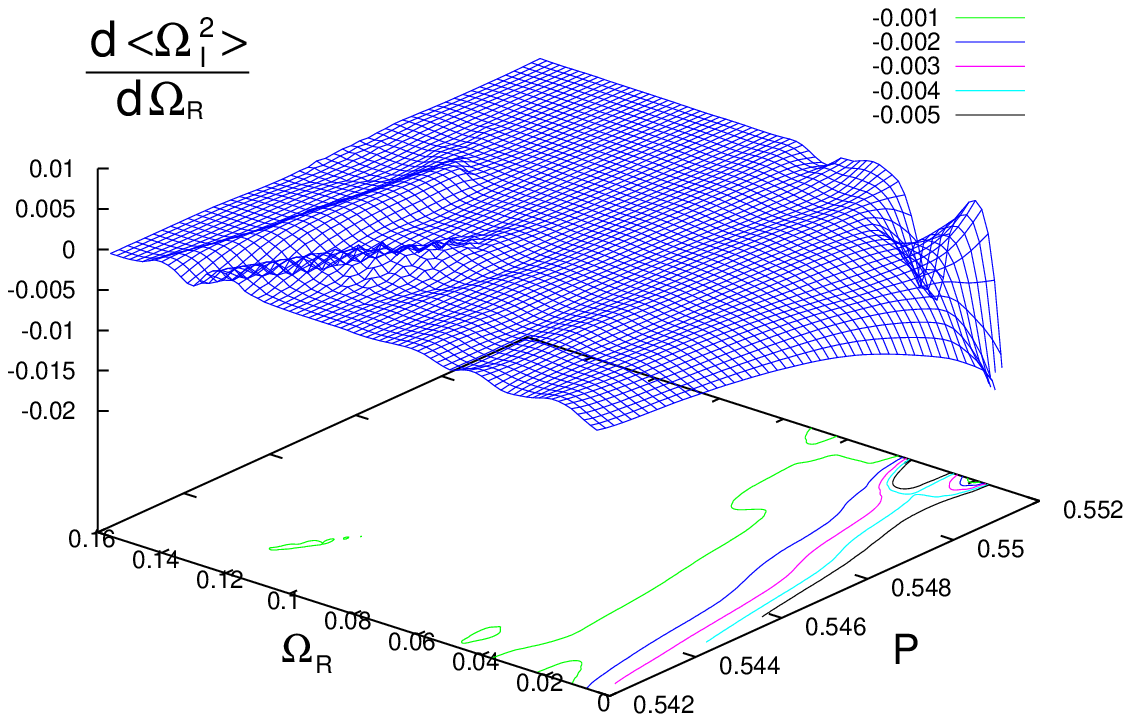} 
}
\caption{$\partial \langle \hat\Omega_{\rm I} ^2 \rangle_{\! c} / \partial P$ (left)
and $\partial \langle \hat\Omega_{\rm I} ^2 \rangle_{\! c} / \partial \Omega_{\rm R}$ (right).}
\label{fig:derOmegaI2}
\end{figure}

We then estimate the effects of the complex phase factor.
In Sec.~\ref{sec:phase}, we have seen that the effects are small in $V_{\rm eff}(\Omega_{\rm R})$ around the critical point of heavy-quark QCD. 
We show that the same is true also for $V_{\rm eff}(P,\Omega_{\rm R})$. 

Applying the cumulant expansion method discussed in Sec.~\ref{sec:phase}, 
the phase factor $ \langle e^{i\hat\theta} \rangle_{\! P, \Omega_{\rm R}} $ in Eq.~(\ref{eq:ppdist}) is expanded as the right-hand side of Eq.~(\ref{eq:cum}),
where 
$\langle \,\cdots\, \rangle_c$ in the present case is the cumulant with fixed $P$ and $\Omega_{\rm R}$.
Because of the symmetry of QCD under $\theta$ to $-\theta$, $\langle \hat\theta^{n} \rangle_c = 0$ for any odd $n$. 
For the case of fixed $\Omega_{\rm R}$, the dominance of the leading $n=2$ term (Gaussian term) was demonstrated by Fig.~\ref{fig:phase} and Appendix B. 
In the same appendix, we also study the application range of the Gaussian approximation with fixing $P$. 
By comparing $\langle \hat\theta^{2} \rangle_c$ and 
$\langle \hat\theta^{4} \rangle_c$ in the expansion of $ \langle e^{i\hat\theta} \rangle_{\! P, \Omega_{\rm R}} $, we find that the Gaussian term dominates in the parameter region near the critical point.
See also Fig.~5 of Ref.~\cite{Saito2} for a further confirmation.

Using the Gaussian dominance, the derivatives of $V_{\rm eff}(P, \Omega_{\rm R}; \beta, \vec\kappa, \vec\mu )$ are given by 
\begin{eqnarray}
   \frac{\partial V_{\rm eff} }{\partial P} 
   &=& \frac{\partial V_0}{\partial P}-6N_{\rm site} \left(\beta^* - \beta_0\right)
          +\frac{\left(3\times 2^{N_t+2} N_s^3 
		  \, q  
		  \right)^2}{2} \frac{\partial \! \left\langle \hat\Omega_{\rm I} ^2\right\rangle_{\! c}}{\partial P} ,
   \label{eq:Veff_mu_P}
\\
   \frac{\partial V_{\rm eff}}{\partial \Omega_{\rm R}} 
   &=& \frac{\partial V_0}{\partial \Omega_{\rm R}}-3\times 2^{N_t+2} N_s^3 \, h 
           +\frac{\left(3\times 2^{N_t+2}N_s^3 
		   \, q  
		   \right)^2}{2} \frac{\partial \! \left\langle \hat\Omega_{\rm I} ^2\right\rangle_{\! c}}{\partial \Omega_{\rm R}} ,
   \label{eq:Veff_mu_Omega}
\end{eqnarray}  
where $V_0(P, \Omega_{\rm R}; \beta) = V_{\rm eff}(P, \Omega_{\rm R}; \beta, \vec0,\vec0 )$
and $q$ is defined by Eq.~(\ref{eq:q}).
When the last terms in these equations modify the curves shown in Fig.~\ref{fig:cont_mu0}, the critical point may shift from that of the phase-quenched theory. 

Our result of $\langle \Omega_{\rm I} ^2 \rangle_c$ is shown in Fig.~\ref{fig:OmegaI2}. 
The statistics is not high around the left and right corners.
We numerically differentiate this data with respect to $P$ and $\Omega_{\rm R}$ using the same method as for the derivatives of $V_0$.
The results of $\partial \langle \Omega_{\rm I} ^2 \rangle_c / \partial P$ and $\partial \langle \Omega_{\rm I} ^2 \rangle_c / \partial \Omega_{\rm R}$ are shown in Fig.~\ref{fig:derOmegaI2}.
From the left panel of this figure, we find that $\partial \langle \Omega_{\rm I} ^2 \rangle_c / \partial P$ is quite flat and small around the critical point
$(P,\Omega_{\rm R}) \approx (0.546,0.06)$. 
According to Eq.~(\ref{eq:Veff_mu_P}), this just causes a small shift of $\beta_{\rm cp}$.
From the right panel, we find that $\partial \langle \Omega_{\rm I} ^2 \rangle_c / \partial \Omega_{\rm R}$ is numerically quite small around the critical point. 
We estimate that, in $N_{\rm f}=2$ QCD, the contribution of the last term in Eq.~(\ref{eq:Veff_mu_Omega}) is at most about 3\% of the second term around the critical point ---
the effect of the complex phase on the critical point is quite small in heavy-quark QCD also in $V_{\rm eff}(P,\Omega_{\rm R};\beta, \vec\kappa,\vec\mu)$. 
Thus, e.g., the critical point in $N_{\rm f}=2$ QCD locates at $\kappa^{N_t} \cosh (\mu/T) = \kappa_{\rm cp}(0)^{N_t} \approx 2\times10^{-5}$ at $\mu\ne0$  as shown in Fig.~\ref{fig:kcp_finitemu}. 
Because $\kappa^{N_t} \sinh (\mu/T) \approx 2\times10^{-5} \times \tanh (\mu/T) < 2\times10^{-5}$ is bounded along the critical curve, the effect of the complex phase factor is under control up to $\mu/T=\infty$.
Similarly, the critical surface in $N_{\rm f} = 2+1$ QCD is given by  Fig.~\ref{fig:crtsur}.

\section{Conclusions}
\label{sec:conclusion}

We have studied the phase structure of QCD in the heavy-quark region by the histogram method. 
When we consider histograms for operators which control the dependence on some coupling parameters in the action, the dependence on these coupling parameters in the histogram becomes in part analytic by the reweighing method.
Because such operators are expected to be sensitive to the phase of the system, we may study the phase structure through a change of the shape of these histograms under a variation of the coupling parameters.
In this paper, we have determined the critical point at which the first-order deconfining transition in the heavy-quark limit turns into a crossover at intermediate quark masses.
We used a histogram for the real part of the Polyakov line, $\Omega_{\rm R}$, as well as that for $\Omega_{\rm R}$ and the plaquette $P$ simultaneously, and compared the results with our previous result obtained using a histogram for $P$ \cite{PRD84_054502}. 
We found that the location of the critical point is consistent among different determinations, implying the robustness of our method to determine the phase structure of the system.

At finite density, the histograms for $\Omega_{\rm R}$ and/or $P$ are factorized into the complex phase factor and phase-quenched part. 
To the leading order of the hopping parameter expansion, the coupling parameter dependence in the phase-quenched part is quite simple for these histograms, and thus the critical point in the phase-quenched theory can be easily computed as a function of the chemical potential $\mu$.
We then estimated the effect of the complex phase factor by the cumulant expansion method.
It turned out that the effect is quite small around the critical point even in the large $\mu$ limit. 
Therefore, the critical point in finite density QCD is almost identical to that for the phase-quenched theory in the heavy-quark region. 
Our results for the critical surface in $N_{\rm f}=2+1$ QCD is shown in Fig.~\ref{fig:crtsur} as functions of the chemical potentials.

As a natural step forward, we are now applying the histogram method to a more realistic case of QCD with light quarks. 
Similar to the case of heavy-quark QCD, the histogram factorizes into the complex phase factor and the phase-quenched part. 
Here, however, unlike the case of heavy-quark QCD, we do expect a significant effect of the complex phase factor such that the crossover at small $\mu/T$ turns into a first-order transition at $\mu/T \sim {\cal O}(1)$. 
A good control of the complex phase factor is essential to clarify the phase structure, getting over the sign problem at $\mu/T > {\cal O}(1)$.
We hope that the cumulant expansion method tested in this paper is helpful in the light-quark region too.
Combining the techniques developed in this paper with Monte Carlo simulations of phase-quenched QCD, 
we are challenging the longstanding issues of finite density QCD.
Our preliminary results look promising \cite{nakagawa}.

\section*{Acknowledgments}
We would like to thank Tetsuo Hatsuda and Yu Maezawa 
for valuable discussions.
This work is in part supported by Grants-in-Aid of the Japanese Ministry of Education, Culture, Sports, Science and Technology (Nos.\ 21340049, 22740168, 20340047, 23540295, and 25287046), 
the Grant-in-Aid for Scientific Research on Innovative Areas (Nos.\ 20105001, 20105003, and 23105706), 
High Energy Accelerator Research Organization (KEK) [No.12/13-14 (FY2012-2013)], 
Center for Computational Sciences (CCS), and Research Center for Nuclear Physics (RCNP).
H.S.\ is supported by the Japan Society for the Promotion of Science for Young Scientists.

\appendix
\section{Effects of the next-to-leading order terms in the hopping parameter expansion}
\label{sec:range}

\begin{figure}[tb]
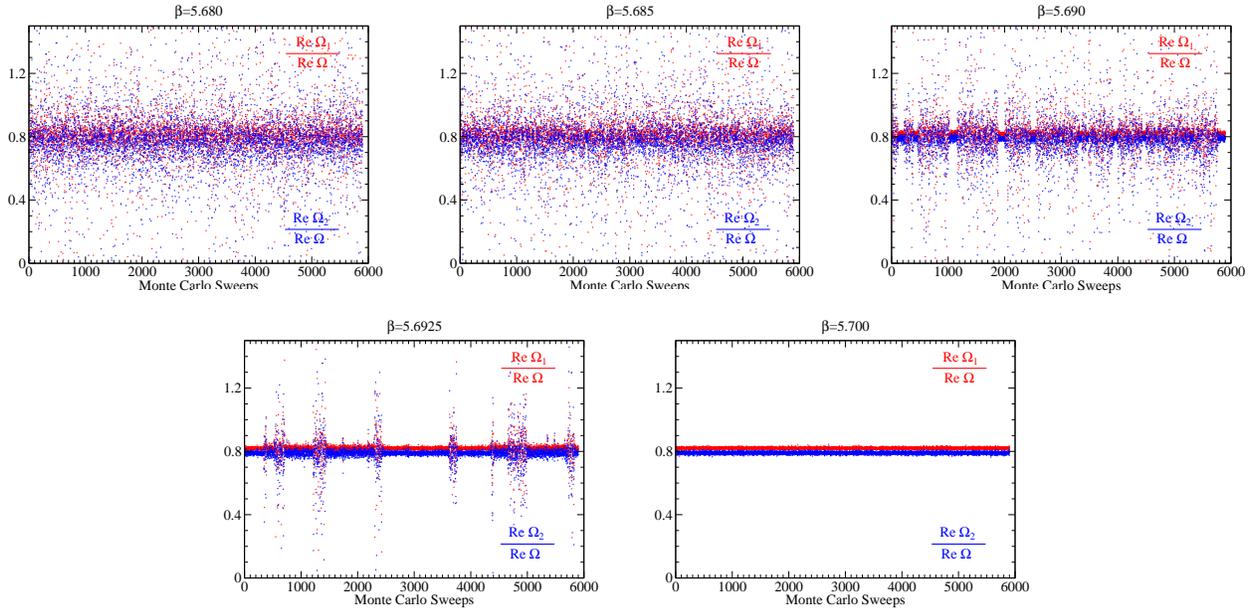

\vspace{5mm}
\centerline{
\includegraphics[width=50mm]{Ratio_Re.NxNt024004.b05680.eps}
\hspace{5mm}
\includegraphics[width=50mm]{Ratio_Re.NxNt024004.b05685.eps}
\hspace{5mm}
\includegraphics[width=50mm]{Ratio_Re.NxNt024004.b05690.eps}
}
\vspace{3mm}
\centerline{
\includegraphics[width=50mm]{Ratio_Re.NxNt024004.b05693.eps}
\hspace{5mm}
\includegraphics[width=50mm]{Ratio_Re.NxNt024004.b05700.eps}
}
\caption{Hstory of ${\rm Re} \Omega_1 / {\rm Re} \Omega$ and ${\rm Re} \Omega_2 / {\rm Re} \Omega$.
The horizontal axis is for the configuration number.
}
\label{fig:timeROmega}
\end{figure}

In this paper, we study the phase structure of heavy-quark QCD using the leading order (LO) of the hopping parameter expansion, Eq.~(\ref{eq:detM}), on an $N_t=4$ lattice. 
In this appendix, we examine the effects from the next-to-leading order (NLO) terms of the hopping parameter expansion, Eq.~(\ref{eq:tayexp}). 

On $N_t=4$ lattices, the NLO terms are $O(\kappa^6)$ and consists of length-6 Wilson loops and length-$N_t+2$ bended Polyakov lines with a handle. 
The former includes rectangular loops, chair-type loops, and parallelogram loops, which are familiar in the improvement program of the lattice gauge action.
Assuming universality of the lattice gauge action, the main effects of these length-6 loops will be to shift the effective gauge coupling and thus will be effectively absorbed by a redefinition of $\beta^*$.

We thus concentrate on the effects of the bended Polyakov lines given by the following expressions:
\begin{eqnarray}
\hat\Omega_1 &\equiv& \frac{1}{6 \cdot4\cdot N^3_s}\sum_{\mathbf{n}}
\sum_{\mu=\pm1}^{3} 
\frac{1}{3} {\rm tr} \left[ \;
U_{\mathbf{n},\mu} U_{\mathbf{n}+\hat\mu,4} U_{\mathbf{n}+\hat4,\mu}^{\dagger} U_{\mathbf{n}+\hat4,4}
U_{\mathbf{n}+2\cdot\hat4,4} U_{\mathbf{n}+3\cdot\hat4,4} \right. \nonumber \\
&& 
+\,U_{\mathbf{n},4} U_{\mathbf{n}+\hat4,\mu} U_{\mathbf{n}+\hat4+\hat\mu,4} U_{\mathbf{n}+2\cdot\hat4,\mu}^{\dagger}
U_{\mathbf{n}+2\cdot\hat4,4} U_{\mathbf{n}+3\cdot\hat4,4} \nonumber \\
&&
+\, U_{\mathbf{n},4} U_{\mathbf{n}+\hat4,4} U_{\mathbf{n}+2\cdot\hat4,\mu} U_{\mathbf{n}+2\cdot\hat4+\hat\mu,4}
U_{\mathbf{n}+3\cdot\hat4,\mu}^{\dagger} U_{\mathbf{n}+3\cdot\hat4,4} \nonumber \\ 
&& \left.
+\,U_{\mathbf{n},4} U_{\mathbf{n}+\hat4,4} U_{\mathbf{n}+2\cdot\hat4,4} U_{\mathbf{n}+3\cdot\hat4,\mu}
U_{\mathbf{n}+3\cdot\hat4+\hat\mu,4} U_{\mathbf{n}+4\cdot\hat4,\mu}^{\dagger}\; \right]
\\
&=&
    \begin{minipage}{10mm}
    \begin{center}
      \vspace{0.4mm}
      \includegraphics[width=8mm]{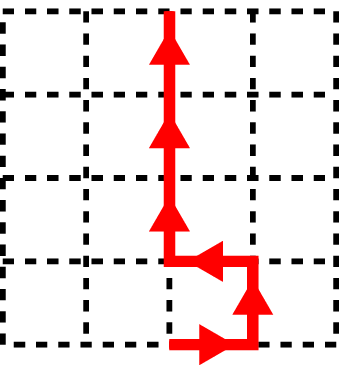}
    \end{center}
    \end{minipage}
    +
    \begin{minipage}{10mm}
    \begin{center}
      \includegraphics[width=8mm]{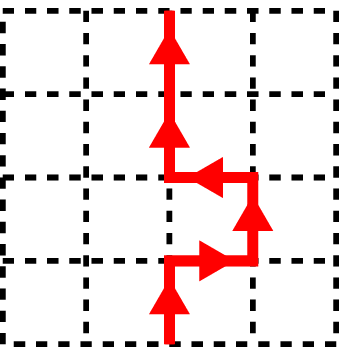}
    \end{center}
    \end{minipage}
    +
    \begin{minipage}{10mm}
    \begin{center}
      \includegraphics[width=8mm]{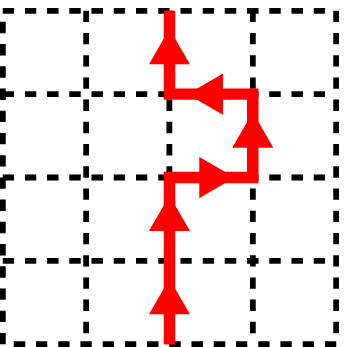}
    \end{center}
    \end{minipage}
    +
    \begin{minipage}{10mm}
    \begin{center}
      \vspace{-0.4mm}
      \includegraphics[width=8mm]{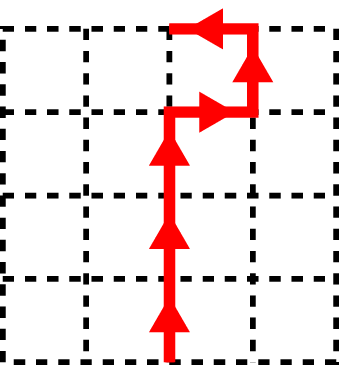}
    \end{center}
    \end{minipage}
\nonumber
\end{eqnarray}
\begin{eqnarray}
\hat\Omega_2 &\equiv& \frac{1}{6\cdot2\cdot N^3_s} \sum_{\mathbf{n}}
\sum_{\mu=\pm1}^{3} 
\frac{1}{3} {\rm tr} \left[ \;
U_{\mathbf{n},\mu} U_{\mathbf{n}+\hat\mu,4} U_{\mathbf{n}+\hat\mu+\hat4,4}
U_{\mathbf{n}+2\cdot\hat4,\mu}^{\dagger} U_{\mathbf{n}+2\cdot\hat4,4} U_{\mathbf{n}+3\cdot\hat4,4} 
\right. \nonumber \\ 
&& \left.
+\,U_{\mathbf{n},4} U_{\mathbf{n}+\hat4,\mu} U_{\mathbf{n}+\hat4+\hat\mu,4} U_{\mathbf{n}+2\cdot\hat4,4}
U_{\mathbf{n}+3\cdot\hat4,\mu}^{\dagger} U_{\mathbf{n}+3\cdot\hat4,4} \; \right]
\\
&=&
    \begin{minipage}{10mm}
    \begin{center}
      \vspace{0.4mm}
      \includegraphics[width=8mm]{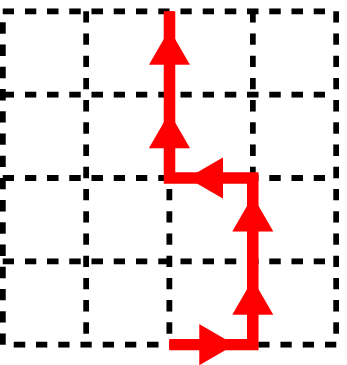}
    \end{center}
    \end{minipage}
    +
    \begin{minipage}{10mm}
    \begin{center}
      \includegraphics[width=8mm]{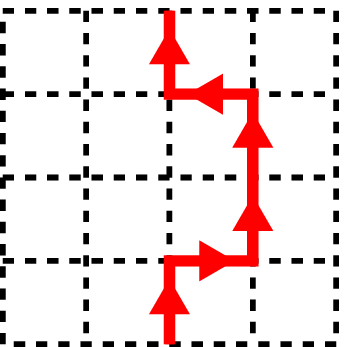}
    \end{center}
    \end{minipage}
\nonumber
\end{eqnarray}
$\hat\Omega_1$ and $\hat\Omega_2$ are normalized such that $\hat\Omega_1 = \hat\Omega_2 =1$ in the weak coupling limit $U_{\mathbf{n},\mu}=1$.
Disregarding the Wilson loop terms, which can be absorbed by $\beta^*$, the quark determinant to the NLO is given by 
\begin{eqnarray}
\ln \left[ \frac{\det M(\kappa,0)}{\det M(0,0)} \right]_{\rm NLO}
&=& 
192 N_s^3 \kappa^4 \left\{
{\rm Re} \hat\Omega+24 \kappa^2 {\rm Re} \hat\Omega_1 +12\kappa^2 {\rm Re} \hat\Omega_2 \right\} .
\label{eq:ln_det_NLO} 
\end{eqnarray}

\begin{table}[t]
\begin{center}
\vspace{5mm}
  \begin{tabular}{cll}\toprule
    $\beta$
    & ${\rm Re} \Omega_1 / {\rm Re} \Omega$
    & ${\rm Re} \Omega_2 / {\rm Re} \Omega$\\ \midrule
    5.6800 &  0.8093(6)   & 0.7756(9) \\
    5.6850 &  0.8108(8)  & 0.7772(10) \\
    5.6900 &  0.8147(6)  & 0.7823(8) \\
    5.6925 &  0.8182(2)  & 0.7871(3) \\
    5.7000 &  0.82029(2) & 0.78992(3) \\ \bottomrule
  \end{tabular}
  \caption{Average of ${\rm Re} \Omega_1 / {\rm Re} \Omega$ and ${\rm Re} \Omega_2 / {\rm Re} \Omega$ for each $\beta$.}
  \label{tab:hpe}
\end{center}
\end{table}

\begin{figure}[t] 
\vspace{7mm}
\centerline{
\includegraphics[width=80mm]{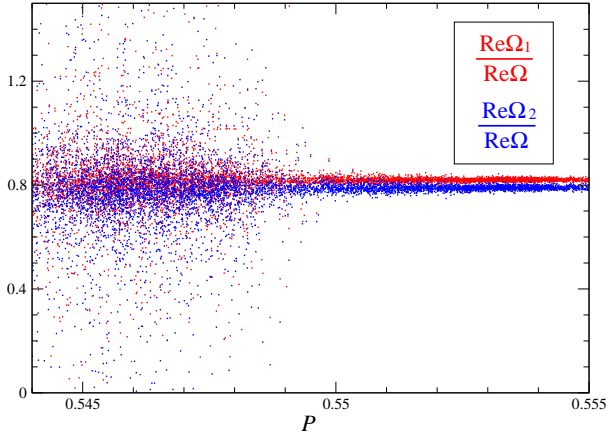} 
}
\caption{${\rm Re} \Omega_1 / {\rm Re} \Omega$ and ${\rm Re} \Omega_2 / {\rm Re} \Omega$ vs. $P$}
\label{fig:ROmega}
\end{figure}

To estimate the effects of the NLO terms, we carry out simulations of the SU(3) Yang-Mills theory on a $16^3 \times 4$ lattice at $\beta=5.68, 5.685, 5.69, 5.6925$, and $5.70$.
With a pseudo heat bath algorithm, we generate 59,000 configurations after 1,000 thermalization sweeps at each $\beta$.
Errors are estimated by a jackknife method with the bin size of 100 sweeps.
Because the branch along the positive real axis of $\Omega$ is relevant at $\kappa>0$, we apply a Z(3) rotation to $\Omega$, $\Omega_1$, and $\Omega_2$ on each configuration such that $\arg\Omega$ is in the range $(-\pi/3,\pi/3)$.

In Fig.~\ref{fig:timeROmega}, we show the history of the ratios ${\rm Re} \Omega_1 / {\rm Re} \Omega$ and ${\rm Re} \Omega_2 / {\rm Re} \Omega$. 
Because the Polyakov lines distribute around $0$ in the low temperature phase, the ratios fluctuate largely at $\beta \simle 5.690$, 
while in the high temperature phase at $\beta \simge 5.700$, the fluctuations are small. 
In the transition region at $\beta\sim 5.690$--$5.6925$, we observe flip-flops between the two phases. 
Besides this $\beta$-dependence of the fluctuation, we find that the central values of 
${\rm Re} \Omega_1 / {\rm Re} \Omega$ and ${\rm Re} \Omega_2 / {\rm Re} \Omega$
are around $0.8$ at all $\beta$.
Averages of the ratios are summarized in Table \ref{tab:hpe}. 
For the calculation of the effective potential $V_{\rm eff}(P)$, the histogram with fixed plaquette value $P$ is important.
In Fig.~\ref{fig:ROmega}, we plot the distribution of the ratios with fixed $P$.
We find that the central values of the ratios are insensitive to $P$.

Assuming ${\rm Re} \Omega_1 / {\rm Re} \Omega \approx  {\rm Re} \Omega_2 / {\rm Re} \Omega \approx  0.8$ independent of $\beta$ and $P$, we may estimate the effect of the NLO terms to the location of the critical point. 
Denoting the values of $\kappa$ for the critical point in LO and NLO as $\kappa_{\rm cp}^{\rm LO}$ and $\kappa_{\rm cp}^{\rm NLO}$, we obtain 
\begin{eqnarray}
192 N_{\rm f} N_s^3 (\kappa_{\rm cp}^{\rm LO})^4 \, {\rm Re} \Omega
\approx 192N_{\rm f} N_s^3 (\kappa_{\rm cp}^{\rm NLO})^4 \,
{\rm Re} \Omega \left[ 1+28.8(\kappa_{\rm cp}^{\rm NLO})^2 \right] .
\end{eqnarray}
Using $\kappa_{\rm cp}^{\rm LO} \approx 0.066$ for $N_{\rm f}=2$, we find $\kappa_{\rm cp}^{\rm NLO} \approx 0.064$; i.e., the effect of the NLO terms in $\kappa_{\rm cp}$ is about 3\% on our $N_t=4$ lattice. 
Alternatively, we may estimate the value of $\kappa$ at which the NLO contributions become comparable to those of the LO terms by 
$
24\kappa^2 ({\rm Re} \Omega_1/{\rm Re} \Omega) +12\kappa^2 ({\rm Re} \Omega_2/{\rm Re} \Omega) 
\sim 1
$. 
Solving this, we find $\kappa \sim 0.18$ for degenerate $N_{\rm f}$ flavor QCD.
Because this is much larger than the critical point $\kappa_{\rm cp} \approx 0.066$ or 0.064,
we conclude that the NLO effects are small around the critical point in heavy-quark QCD at $N_t=4$.

\section{Application range of the Gaussian approximation for the complex phase distribution}
\label{sec:gaussap}

\begin{figure}[t] 
\centerline{
\includegraphics[width=8.5cm]{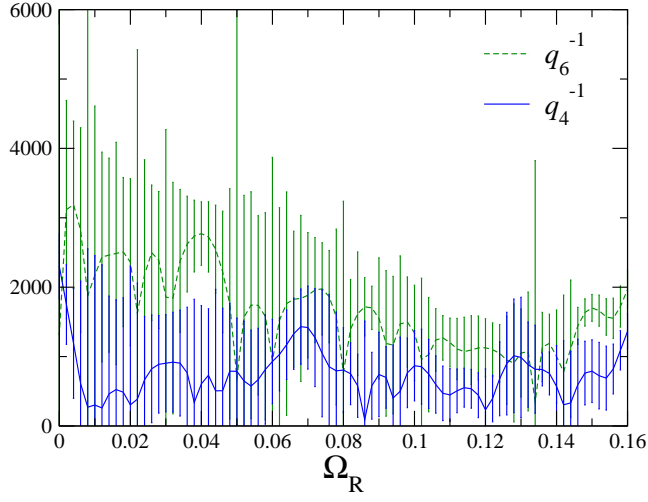}
}
\vspace{-2mm}
\caption{$q_4^{-1}$ (blue)  and $q_6^{-1}$ (green) as a function of $\Omega_R$.}
\label{fig:conv}
\end{figure}

\begin{figure}[t] 
\centerline{
\includegraphics[width=8.5cm]{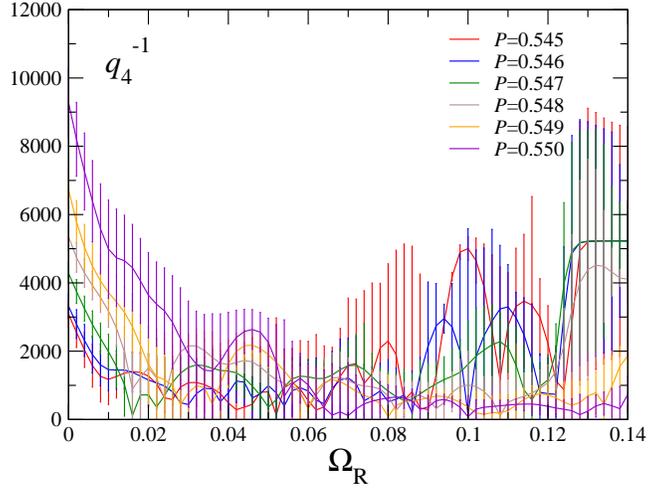}
}
\vspace{-2mm}
\caption{$q_4^{-1}$ as a function of $P$ and $\Omega_R$.}
\label{fig:convPO}
\end{figure}

We estimate the range of coupling parameters where higher order terms in the cumulant expansion are smaller than the leading term in the complex phase distribution in heavy-quark QCD. 
The cumulant expansion is introduced in Sec.~\ref{sec:phase} as
\begin{eqnarray}
\left\langle e^{i \hat\theta} \right\rangle_{\Omega_{\rm R}; (\beta^*, \vec0,\vec0)} 
= \exp \left[ \sum_{n=1}^{\infty} (-1)^n \frac{\langle \hat\theta^{2n} \rangle_c}{(2n)!} \right], 
\label{eq:cuma}
\end{eqnarray}
where the complex phase $\hat\theta$ is given by Eq.~(\ref{eq:thetaOmegaI}) in terms of the imaginary part of the Polyakov line $\hat\Omega_{\rm I}$ 
and 
$\left\langle \hat{\theta}^{2n} \right\rangle_c $
is given by Eq.~(\ref{eq:theta2nc}) 
with $q = \sum_{f=1}^{N_{\rm f}} \kappa_f^{N_t} \sinh\left( \frac{ \mu_f}{T} \right)$.
Thus the ratio $\langle \hat{\theta}^{2n} \rangle_c / \langle \hat{\theta}^{2} \rangle_c$ is given by 
\begin{eqnarray} 
\frac{\left\langle \hat{\theta}^{2n} \right\rangle_c}{\left\langle \hat{\theta}^{2} \right\rangle_c} = \left( 3\times 2^{N_t+2} N_s^3 
\, q  
\right)^{2(n-1)} 
\frac{\left\langle \hat\Omega_{\rm I}^{2n} \right\rangle_c}{\left\langle \hat\Omega_{\rm I}^{2} \right\rangle_c}
\end{eqnarray} 
for $n=2, 3, 4, \cdots$.
Let us define $q_{2n}$ as the value of $q$ at which 
$| \langle \hat{\theta}^{2n} \rangle_c |/(2n)! = | \langle \hat{\theta}^{2} \rangle_c |/2!$:
\begin{eqnarray} 
q_{2n}^{-1}
= 3 \times 2^{N_t+2} N_s^3 
\left| \frac{2 \left\langle \hat\Omega_{\rm I}^{2n} \right\rangle_c}{(2n)! \left\langle \hat\Omega_{\rm I}^{2} \right\rangle_c} \right|^{\frac{1}{2(n-1)}}.
\end{eqnarray} 

We first study the case discussed in Sec.~\ref{sec:poldis}, i.e., the $\hat\theta$ distribution with fixing $\Omega_R$.
We plot $q_{2n}^{-1}$ as a function of $\Omega_R$ in Fig.~\ref{fig:conv}.
The solid blue line is for $q_4^{-1}$ and the dashed green line for $q_6^{-1}$. 
We find that $q_4^{-1}$ and $q_6^{-1}$ are smaller than about 3000 in the whole range of $\Omega_R$.
This means that $| \langle \hat{\theta}^{4} \rangle_c |/4!$ and 
$| \langle \hat{\theta}^{6} \rangle_c |/6!$ are much smaller than 
$| \langle \hat{\theta}^{2} \rangle_c |/2!$ when 
$q = \sum_{f=1}^{N_{\rm f}} \kappa_f^{N_t} \sinh (\mu_f/T) \ll 3 \times 10^{-4}$.
This condition is well satisfied at 
$q= N_{\rm f} \kappa^{N_t} \sinh (\mu/T) = 4 \times 10^{-5}$ 
and $1 \times 10^{-4}$ in $N_{\rm f}=2$ QCD discussed in Sec.~\ref{sec:poldis}.
In fact, $\langle \hat{\theta}^{2} \rangle_c /2!$ dominates over 
$\langle \hat{\theta}^{4} \rangle_c /4!$ and $\langle \hat{\theta}^{6} \rangle_c /6!$ 
at these points as shown in Fig.~\ref{fig:phase}.
This suggests that the effects from the higher order cumulants are small at these points, as explicitly demonstrated in Fig.~\ref{fig:phase} by the agreement of the second order result (black dashed lines) and the full result (red solid lines).

We then study the case with fixing $P$ and $\Omega_R$.simultaneously, as studied in Sec.~\ref{sec:popot}.
In Fig.~\ref{fig:convPO}, $q_{4}^{-1}$ is shown as a function of $P$ and $\Omega_R$.
We find that $q_{4}^{-1}$ is small, except for the regions of small $P$ and large $\Omega_R$ and large $P$ and small $\Omega_R$, 
i.e. the top-left and bottom-right corners in Fig.~\ref{fig:contdvdpdO}, 
at which statistics is not sufficiently high.


\end{document}